\documentclass[amsmath,amssymb,twocolumn,superscriptaddress]{revtex4}

\usepackage{graphicx}
\usepackage{dcolumn}

\def\be{\begin{equation}}
\def\ee{\end{equation}}
\def\beq{\begin{eqnarray}}
\def\eeq{\end{eqnarray}}

\usepackage{bm}
\usepackage{graphicx}
\usepackage{dcolumn}
\usepackage{amsmath}
\usepackage[latin1]{inputenc}
\usepackage{graphicx, psfrag}
\usepackage{amssymb}
\usepackage[colorlinks=true, citecolor=blue, urlcolor = blue, linkcolor= red, bookmarks=true]{hyperref}
\usepackage{float}
\usepackage{amsmath}
\usepackage{amsfonts}
\usepackage{dcolumn}
\usepackage{hyperref}
\usepackage{subfigure}
\usepackage{pgfplots}
\usepackage{epstopdf}
\usepackage{booktabs}

\begin{document}


\title{Role of gravitational decoupling on isotropization and complexity of self-gravitating system under complete geometric deformation approach}



\author{S. K. Maurya}%
\email[Email:]{sunil@unizwa.edu.om}
\affiliation{Department of Mathematical and Physical Sciences,
College of Arts and Sciences, University of Nizwa, Nizwa, Sultanate of Oman}

\author{Riju Nag} \email[]{rijunag@gmail.com } 

\affiliation{Department of Mathematical and Physical Sciences,
College of Arts and Sciences, University of Nizwa, Nizwa, Sultanate of Omana}


\date{\today}

\begin{abstract}
In the present paper, we discuss the role of gravitational decoupling to isotropize the anisotropic solution of Einstein's field equations in the context of the complete geometric deformation (CGD) approach and its influence on the complexity factor introduced by L. Herrera (Phys. Rev. D 97, 044010 (2018)) in the static self-gravitating system. Moreover, we also proposed a simple and effective technique to generate new solutions for self-gravitating objects via CGD approach by using two systems with the same complexity factor and vanishing complexity factor proposed by Casadio et al. Eur. Phys. J. C 79, 826 (2019). The effect of decoupling constant and the compactness on the complexity factor have been also analyzed for the obtained solutions. 
\end{abstract}


\maketitle


\section{Introduction}\label{sec1}
Quantifying the term "complexity" has been quite a fascinating challenge among researchers. Depending on various physical problems, the term complexity changes its notion. For example, a perfect crystal can be thought of as a system with zero complexity. Here, zero complexity means the crystal structure is perfectly ordered and periodic. In contrast, an isolated ideal gas is fully disordered and it contains maximum information as the system can be obtained in any accessible state with equal probability. Now, if we consider the concept of "disequilibrium" i.e. how a system deviates from the equilibrium, we find that the ideal gas has minimum disequilibrium while the perfect crystal has maximum disequilibrium. So the contrasting views of complexity while considering "information" and "disequilibrium" can be addressed by defining complexity as a product of these concepts \cite{c7}. So in this way, complexity is zero for both perfect crystal and ideal gas, as it should be. The work of Lopez-Ruiz and collaborators \cite{c7, c10} about complexity has been extended to self gravitating systems \cite{c13,c14,c15,c16,c17,c18}. Recently, Carrasco-Hidalho and Contreras \cite{c20} proposed a polynomial complexity factor containing gravastar model \cite{c19} as a special case.  Contreras and Fuenmayor  \cite{c21} considered complexity factor as a physical quantity containing anisotropy and gradients in the density. L. Herrera \cite{c22} proposed a new definition of complexity for self-gravitating, spherically symmetric systems, based on a specific parameter that arises in the orthogonal splitting of Riemann tensor. Some more recent works in this regard can be found in \cite{c23,c24,c25,c26}. \\
It can be said, that fixing some value of the complexity factor for a specific scenario (example: a system having vanishing complexity) can act as an equation of state which may lead close approximation to Einstein's field equations. But for these equations, it is very difficult to obtain analytical solutions. In this scenario, a recent well-known tool called Gravitational Decoupling (GD) by means of the minimal geometric deformation (MGD) \cite{o1} and its extension known as complete geometric deformation \cite{m61} formalism works perfectly to convert the isotropic solutions into anisotropic domain or can be even used to obtain new solutions. Moreover, it is well-known that the MGD tool is a transformation that is performed on the metric potential along with the radial component of the line element by introducing a decoupler function. As a consequence, the original system splits into two relatively simpler sets of the equations. Another great advantage of MGD is that it can extend a simple solution to more generalized and complex cases by the addition of an extra source ($\Theta^i_j$) with the original energy-momentum tensor via coupled with a dimensionless parameter. Later on, Ovalle and his collaborators found a drawback in the MGD approach, such as considering only radial transformation can't explain a stable black hole with a well-defined horizon. In this regard, the MGD was extended to deform both radial and temporal metric functions \cite{m61}. Mathematically, in the extended case of MGD, the deformation acts in the following way:
\begin{eqnarray}
\nu(r) \mapsto \xi(r) + \alpha h(r),~~~~ \text{and}~~~~ \lambda(r) \mapsto - \ln [\mu(r) + \alpha f(r)]\nonumber
\end{eqnarray}
It is noted here that the extended gravitational decoupling (EGD) demands the supposition of a seed solution, which allows reducing the number of free variables (or number of degrees of freedom).  Due to this, we need only two extra conditions in order to close the system. There are many ways to solve the equation despite the appearance of the new degree of freedom $h(r)$ in the $\theta$-sector, such as implementing the mimick approach with particular form of $h(r)$, or using EoS approach together with the mimick approach \cite{sunilepjc1,sunilepjc2,sunilepjc3} to solve the $\Theta$-sector. On the other hand, it must be noted that the hydrostatic balance gets severely modified because of the deformation. Therefore, it is very important to check the hydrostatic balance in order to assess the viability of the solution.  Recently some interior solutions were generalized into the anisotropic domain using gravitational decoupling via MGD and CGD approaches in different contexts \cite{MGD0,MGD1,MGD2,MGD3,MGD4,MGD5,MGD6,MGD7,MGD8,MGD9,MGD10,MGD11,MGD12,MGD13,MGD14,MGD15,MGD16,MGD17,MGD18,MGD21,MGD22}. 
Often for some cases like extremely dense compact star solutions, the anisotropic scenario gives more realistic results, whereas, in some other cases, the isotropic solution is more necessary. In compact stars, due to extreme internal density and strong gravity, the pressure components break down into two components, i.e. radial and tangential pressure. For the matter to be isotropic in nature, these two components of pressure have to be of the same value. In this connection, Ruderman \cite{rud} showed that for densities higher than $10^{15} g/cm^3$, the two components of pressure don't have the same magnitude and nuclear matter transforms into anisotropic distribution. \\ 
On the other hand, the finding of the new physical viable anisotropic solution for a static self-gravitating system is easier than the isotropic solution of Einstein's field equations. Till now, the researchers have obtained around 130 interior solutions of Einstein's field equations (EFE) for perfect fluid matter distributions but only few of them are well-behaved that can be used for modeling of the self-gravitating compact objects \cite{Delgaty}. This is why it is still a challenge to obtain the new well-behaved isotropic solution of the Einstein field equations. Therefore, recently Casadio and his collaborators \cite{m79} have proposed is a very powerful methodology, known as isotropization techniques via gravitational decoupling using the MGD approach to find the new isotropic solutions for any known seed spacetime geometry corresponding to anisotropic matter distribution. In this work, they have investigated a new isotropic solution using the MGD approach as well as two other gravitationally decoupled anisotropic solutions corresponding to two systems with the same complexity factor, and zero complexity factor using Tolman IV solution. Some recent solutions on anisotropic star with different complexity factor can be seen in the following works \cite{GDC1,GDC2,GDC3,GDC4,GDC5}. \\ 
 In the current article, we develop an isotropization technique using gravitational decoupling in the framework of a complete geometric deformation (CGD) approach to find the new isotropic solutions from a known spacetime geometry for the anisotropic matter distribution. An example has been presented to validate this methodology. Moreover, we also discussed the complexity factor and the effect of the decoupling constant on complexity factor for this isotropic solution. The present simple methodology has also been utilized to obtain the new anisotropic solutions by taking two systems with the same complexity factor as well as for zero complexity factor using the Karori-Barua solution.
 
 The article is arranged as follows: The section I is the introduction, while in the section II, the Einstein field equation for two sources by gravitational decoupling has been discussed. The  section III consists of the method of isotropization of the gravitationally decoupled system and new solution obtained by taking Tolman-Kuchowicz spacetime for seed solution. The complexity by gravitational decoupling has been analyzed in the section IV. In this section, we also discussed the complexity factor for isotropic solution obtained in section III. In section V, we have investigated new anisotropic EGD solutions generated by two systems with same complexity factors and vanishing complexity factor using Karori-Barua seed solution, which are presented in subsections A and B. The last section VI contains the discussions and conclusions of the article.   
 
\section{Einstein's field equation for two sources introduced by gravitational decoupling }
We propose the brief review of Einstein's field equations with two different sources, 
\begin{eqnarray} \label{eq1}
 &&   R_{ij}-\frac{1}{2}\,g_{ij}\,R= -8\pi(T_{ij}+\beta\,\Theta_{ij})
\end{eqnarray}
Now with $G = c = 1$, the relativistic units are considered to express the field equations for the Ricci tensor denoted by $R_{ij}$, and $R$ is contracted Ricci scalars, and $\beta$ is decoupling constant. 
Here, $T_{ij}$ denote the energy-momentum tensor and the source $\theta_{ij}$ may contain new fields, like scalar, vector and tensor fields. 
Since the Einstein tensor (\ref{eq1}) satisfy the Bianchi identity, therefore the effective energy momentum tensor $\hat{T}_{ij}=T_{ij}+\beta\,\Theta_{ij}$  must be conserved, that is,
\begin{eqnarray}
&& {\nabla_i}\hat{T}^{ij}=0, \label{eq2}
\end{eqnarray}
The following static spherically symmetric line element is taken for describing the space-time of interior region of the stellar system as, 
\begin{eqnarray} \label{eq3}
 ds^2 = -e^ {\lambda (r)} dr^2 - r^2\big(d\theta^2 + \sin^2 \theta~ d\phi^2\big)~+ e^{\nu( r)} dt^2 ,
\end{eqnarray}
 where the metric potentials $\nu$ and $\lambda$ are only radially dependent. However, the effective energy momentum tensor $\hat{T}_{ij}$ is considered for anisotropic matter distribution,
\begin{eqnarray}
&& \hat{T}_{ij} = \left({\epsilon}+{P}_r\right)u_{i}u_{j}+{P}_\perp g_{ij}+({P}_r-{P}_\perp)\chi_{i}\,\chi_{j},\label{eq4}
\end{eqnarray}
where ${P}_r$ and ${P}_\perp$ denote the radial and tangential pressures, respectively while ${\epsilon}$ is the energy density of matter.\ Moreover, $u^{i}$ denotes a contravariant $4$-velocity and $\chi^i=\sqrt{1/g_{rr}}\,\delta^i_1$ is a unit space-like vector in the radial direction.
Then under the line element (\ref{eq3}), the Einstein field equation (\ref{eq1}) with Eq.(\ref{eq4}) provides the following differential equations, 
\begin{eqnarray}
\label{eq5}
&&\hspace{-0.6cm} 8\pi {\epsilon}=T^0_0+\beta\,\Theta^0_0=\frac{1}{r^2}-e^{-\lambda}\left(\frac{1}{r^2}-\frac{\lambda^{\prime}}{r}\right),
\\ 
\label{eq6}
&&\hspace{-0.6cm} 8\pi {P}_{r}=-T^1_1-\beta\,\Theta^1_1=-\frac{1}{r^2}+e^{-\lambda}\left(\frac{1}{r^2}+\frac{\nu^{\prime}}{r}\right),
\\
\label{eq7}
&&\hspace{-0.6cm} 8\pi {P}_{\perp}=-T^2_2-\beta\,\Theta^2_2=\frac{e^{-\lambda}}{4}\left(2\nu^{\prime\prime}+\nu^{\prime2}-\lambda^{\prime}\nu^{\prime}+2\frac{\nu^{\prime}-\lambda^{\prime}}{r}\right),~~~
\end{eqnarray}
and, the conservation equation for system (\ref{eq5}) - (\ref{eq7}) will become,
\begin{eqnarray}\label{eq8}
&&\hspace{-0.5cm} (P_r)^\prime + \frac{\nu^\prime}{2} \big( \epsilon + P_r \big) - \frac{2}{r}(P_\perp-P_r) =  -(T^1_1)^{\prime} \nonumber\\ && + \frac{\nu^{\prime}}{2} \left(T^0_0 - T^1_1\right) +\frac{2(T^2_2-T^1_1)}{r}-  \beta\, L(\Theta_i^i) =0, ~~
\end{eqnarray}
where the function $L(\theta_i^i)$ is given by 
\begin{equation} \label{eq9}
    L(\Theta_i^i) \equiv  \left(\Theta^{1}_{1}
    \right)^{\prime}  + \frac{\nu^{\prime}}{2} \left(\Theta^{0}_{0}-\Theta^{1}_{1} \right) + \frac{2}{r} \left(\Theta^{1}_{1}-\Theta^{2}_{2}\right).
\end{equation}
Now it is important to mention here that the source $T_{ij}$ can describe perfect fluid or anisotropic fluid matter distribution. Suppose It describes an anisotropic fluid matter distribution then, the effective density and effective pressures can be read as, 
\begin{eqnarray} \label{eq10}
 && \epsilon=  T^0_0+\beta\,\Theta^0_0 =\rho+\beta\,\Theta^0_0, \\ 
 && P_r=  -T^1_1-\beta\,\Theta^1_1,=p_r-\beta\,\Theta^1_1~ \label{eq11} \\
&& P_\perp=  -T^2_2-\beta\,\Theta^2_2=p_t-\beta\,\Theta^2_2. \label{eq12} 
\end{eqnarray}
 where, $\rho$, $p_r$ and $p_t$ denote the energy density, radial pressure and tangential pressure, respectively.  Then the effective anisotropy can be given as,
 \begin{eqnarray} \label{eq13}
 \hat{\Pi}=P_\perp-P_r=\Pi+\beta\, \Pi_{\Theta}
 \end{eqnarray}
where, 
\begin{eqnarray}
&& \Pi=p_t-p_r~~~ \text{and}~~~ \Pi_{\Theta}=(\Theta^1_1-\Theta^2_2). \label{eq14} 
\end{eqnarray}
Here the anisotropy $\Pi_{\Theta}$ is generated by second source $\Theta_{ij}$, and Misner-Sharp mass function $m(r)$ for the effective system can be calculated by the formula,
\begin{eqnarray}\label{eq15}
 m(r) &=& \frac{r}{2} [1-e^{-\lambda(r)}]=4\pi \int^r_0 x^2 \epsilon(x) dx  \nonumber\\ &=& \underbrace{4\pi \int^r_0 x^2 \rho(x) dx}_{m_{GR}}+ \underbrace{4\pi \beta \int^r_0 x^2 \Theta^0_0(x) dx}_{m_{\Theta}}.~~~~~~~
\end{eqnarray}
The $m_{GR}$ and $m_{\Theta}$ represent the mass function due matter distribution $T_{ij}$ and $\Theta_{ij}$, respectively.\\
Also, there is an another definition in order describe the energy content inside a fluid sphere which was proposed by Tolman many years before. The Tolman mass ($m_T$) for the spherically symmetric static
spacetime (\ref{eq3}) and energy-momentum tensor $\hat{T}_{ij}$ can be given by the formula [19]
\begin{eqnarray} \label{eq16}
m_T=4\pi\,\int^r_0 x^2\,e^{(\lambda+\nu)/2} (\rho+P_r+2P_\perp) dx,
\end{eqnarray}
The above formula was proposed in order a measure of the energy contained inside a fluid sphere of radius $r$. However, Tolman mass function $m_T$ using the field equations (\ref{eq5})-(\ref{eq7}) under the spacetime (\ref{eq3}) can be written as,
\begin{eqnarray}
m_T=\frac{r^2\,\nu^\prime}{2}\,e^{(\nu-\lambda)/2}, \label{eq17}
\end{eqnarray}
The above formula states about the physical meaning of $m_T$ 
as the active gravitational mass. Since instantaneously at rest in a static gravitational field, the gravitational acceleration of a test particle is given by (see \cite{c22}, for more details)
\begin{eqnarray}
a=\frac{\nu^\prime\,e^{-\lambda}}{2}=\frac{e^{-\nu/2}\, m_T}{r^2}
\end{eqnarray}
Now we apply the extended gravitational decoupling by means of a complete geometric deformation (CGD) approach in order to see the general effects of the extra source $\Theta$ on the energy-momentum tensor $T_{ij}$. Under this, the metric functions $e^{\lambda}$ and $e^{\nu}$ undergone by the following transformation  Ovalle \cite{m61} as, 
\begin{eqnarray}\label{eq19}
    &&  \xi(r) \mapsto \nu(r) = \xi(r) + \alpha ~ h ( r ), \\\label{eq20}
    &&  \mu(r) \mapsto e^{-\lambda(r)} = \mu(r) + \alpha ~ f (r ).
\end{eqnarray}
where, $f(r)$ and $h(r)$ denote the geometric deformation functions for the radial and temporal metric components, respectively. Since here we are considering the extended case therefore we need to set $f(r)\ne 0$ and $h(r)\ne 0$. Then the transformations (\ref{eq19}) and (\ref{eq20}) allow us to split the  field equations (\ref{eq5})-(\ref{eq7}) into two sets of equations: (i) the standard Einstein field equations corresponding to energy-momentum tensor $T_{ij}$ (same as at $\alpha$ = 0) as
\begin{eqnarray}\label{eq21}
 && 8\pi {\rho}= \frac{1-\mu}{r^2} - \frac{\mu^ \prime}{r} ,\\\label{eq22}
&&  8\pi p_r=\frac{\mu -1 }{r^2} - \frac{\mu ~~ \xi ^ \prime}{r}  ,\\\label{eq23}
&&   8\pi p_t= \mu \bigg( \frac{\xi ^{\prime \prime}}{2} + \frac{\xi ^{ \prime 2}}{4} + \frac{\xi^\prime}{2r} \bigg) + \bigg( \frac{\xi ^ \prime \mu^\prime}{4} + \frac{\mu^\prime}{2r}\bigg),~~~
\end{eqnarray}
with the conservation equation,
\begin{eqnarray}
(p_r)^\prime + \frac{\xi^\prime}{2} \big( \rho + p_r \big) = \frac{2\Pi}{r}. \label{eq24}
\end{eqnarray}
and the solution of this system can be described by the following spacetime,
\begin{eqnarray}\label{eq25}
  ds^2  = -\mu dr^2 - r^2\big( d\theta^2 - \sin^2 \theta d \phi^2
 \big) + e^ \xi dt^2, 
\end{eqnarray}
 with
 \begin{eqnarray}
 \mu=1-\frac{2m_{GR}}{r}=4\pi \int^r_0 x^2 \rho(x) dx. \label{eq26}
\end{eqnarray}
(ii) Now the second set of equations for the extra source is determined by turning on $\beta$ as, 
\begin{eqnarray}\label{eq27}
&& 8\pi \Theta^0_0 =- ~~\bigg(\frac{f^\prime}{r} + \frac{f}{r^2}\bigg),\\\label{eq28}
&& 8\pi\Theta^1_1 = - f ~~\bigg( \frac{\nu^\prime}{r} + \frac{1}{r^2} \bigg) -  \frac{\mu h^\prime }{r} ,\\\label{eq29}
&&8\pi\Theta^2_2= - \frac{ f }{2} ~~\bigg( \nu ^{\prime\prime} + \frac{\nu^{\prime2}}{2} + \frac{\nu^\prime}{r} \bigg) - \frac{\alpha f^\prime }{2} ~~ \bigg(\frac{\nu^\prime }{2} + \frac{1}{r}\bigg) \nonumber\\&&\hspace{1.2cm}-  \frac{\mu}{4}\big(2 h^{\prime\prime} + \alpha h^{\prime2} + \frac{2~h^\prime}{r} + 2 \xi^\prime h^\prime\big) - \frac{\mu^\prime ~h^\prime}{4}.~~~~~~
 \end{eqnarray}
 whose conservation equation read as, 
\begin{eqnarray} \label{eq30}
&& -\frac{\nu^\prime }{2}\big(\theta^0_0-\theta^1_1\big) +\big(\theta^1_1 \big)^\prime -\frac{h^\prime }{2} \big(p_r + \rho\big)=\frac{2\Pi_{\Theta}}{r}. ~~~~~~~~
\end{eqnarray}
\section{Isotropization of gravitationally decoupled system}\label{sec3}
In this section, we will adopt the systematic approach proposed by Casadio and his collaborators \cite{m79} to isotropize the decoupled system (\ref{eq5})-(\ref{eq7}) under EGD scenario.  As we discussed previously, the effective anisotropy $\hat{\Pi}$ given by (\ref{eq13}) may not be same as the anisotropy $\Pi$ due to extra contribution $\beta\,\Pi_{\Theta}$. Here, our aim is to isotropize the effective system, which can be obtained by setting $\hat{\Pi}=0$  with assuming $\Pi\ne0$ [see Ref.\cite{m79} for more details]. Therefore, $\hat{\Pi}=\Pi+\beta\,\Pi_{\Theta}=0$ leads 
\begin{eqnarray}
\Pi=-\beta\,\Pi_{\Theta}~~\Longrightarrow~~ \Pi=-\beta(\Theta^1_1-\Theta^2_2). \label{eq31}
\end{eqnarray}

Now by plugging the Eqs.(\ref{eq28}) and (\ref{eq29}) in Eq.(\ref{eq31}), we get the following non-linear differential equation,
\begin{eqnarray}
 f^\prime (2 + \nu^\prime r)+f (-4 - 2 \nu^\prime r + 2 \nu^{\prime \prime} r^2 + \nu^{\prime2} r^2) + 
 r (-2 h^\prime \mu \nonumber\\ + h^\prime \mu^\prime r + 2 h^{\prime \prime} \mu r + b h^{\prime2} \mu r + 2 h^\prime \nu^\prime \mu r) + \frac{4r^2}{\beta}\,\Pi=0.~~~\label{eq32}
\end{eqnarray}
As we can see that the above equation (\ref{eq32}) is a first order linear ODE in $f(r)$ while it is a second order non-linear in $h(r)$.  Therefore, we solve the above differential for $f(r)$ due to simplicity. Now we assume a spacetime corresponding energy-momentum tensor $T_{ij}$ generated by Tolman-Kuchowicz  metric functions $\{\xi,\,\mu\}$ 
\begin{eqnarray}
 ds^2=-(1+Kr^2+Lr^4)\, dr^2- r^2 d\Omega_2^2 ~+ e^{Ar^2+B} dt^2.~~~\label{eq33}
\end{eqnarray}
 together with temporal deformation function $h(r)=Cr^2$ in order to isotropize the gravitationally decoupled system (\ref{eq5})-(\ref{eq7}). Then, the metric functions $\mu(r)=1/(1+Kr^2+Lr^4)$ and $e^{\xi(r)}=e^{Ar^2+B}$ describe the anisotropic solution for the system (\ref{eq21})-(\ref{eq23}). The constant parameters $K$, $L$, $A$, and $B$ will be determined by matching of the seed spacetime (\ref{eq33}) with exterior vacuum spacetime at surface $r=R$, If we consider exterior vacuum spacetime is described by exterior Schwarzschild solution, then
\begin{eqnarray}
&& 1-\frac{2M_s}{R}=e^{\xi(R)},~~~\label{eq34}\\
&& 1-\frac{2M_s}{R}=\mu(R),~~~\label{eq35}\\
&& p_r(R)=0,~~~\label{eq36}
\end{eqnarray}
where $m_{GR}(R)=M_s$ is the total mass of the seed spacetime (\ref{eq33}) related to the energy-momentum tensor $T_{ij}$. Using the conditions (\ref{eq34})-(\ref{eq36}), we find the constants $A$, $B$, and $K$
\begin{eqnarray}
&& A=\frac{M_s}{R^2 (R-2 M_s)},\\
&& B=\frac{-4 M_s^2+5 M_s R-R^2}{(2 M_s-R) R},\\
&& K=\frac{-2 M_s-2 L M_s R^4+L R^5}{(2 M_s-R) R^2}.
\end{eqnarray}
By plugging of spacetime geometry (\ref{eq33}) into Eq.(\ref{eq32}) and using $h=Cr^2$, we obtain the deformation  function $f(r)$ as
\begin{widetext}
\begin{eqnarray}
&&\hspace{-0.6cm}f(r)= \frac{r^2 e^{-\left(r^2 (A+b C)\right)}}{\beta \sqrt{K^2-4 L} \left(K r^2+L r^4+1\right) \left(A^2+2 A \beta C-A K+\beta^2 C^2-\beta C K+L\right)} \Bigg[e^{-\frac{A K+\beta C K+L}{L}} \Bigg(e \Bigg\{\sqrt{K^2-4 L} \left(K+L r^2\right)\nonumber\\&&\hspace{0.0cm}\times \big(A^2+2 A \beta C-A K+\beta^2 C^2-\beta C K+L\big) e^{\frac{(A+\beta C) \left(K+L r^2\right)}{L}}+\beta^2 C^2 f_2(r) \left(K r^2+L r^4+1\right)  \chi_1 +\beta^2 C^2 f_3(r)\, \big(K r^2+L r^4\nonumber\\&&\hspace{0.0cm} +1\big) \chi_2 \Bigg\}-f_1(r) \sqrt{K^2-4 L} (A+\beta C) \left(K r^2+L r^4+1\right) e^{\frac{K (A+\beta C)}{L}} \left(A^2+2 A \beta C-A K+3 \beta^2 C^2-\beta C K+L\right)\Bigg)+F \Bigg],~~~~~\label{eq40}
\end{eqnarray}
where, $F$ is a constant of integration, and then the solution of the system (\ref{eq5})-(\ref{eq7}) can be described by the spacetime  
\begin{eqnarray}
 ds^2=-\frac{(1+Kr^2+Lr^4)}{1+\beta\,(1+Kr^2+Lr^4)\,f(r)} dr^2- r^2 d\Omega_2^2 + \big(e^{Ar^2+B+\beta Cr^2}\big) dt^2. \label{eq41}
\end{eqnarray}
However, the effective energy density and effective pressures can be given as,
\begin{small}
\begin{eqnarray}
&&\hspace{-0.5cm}P_r(r,\,\beta)= \frac{1}{K r^4+L r^6+r^2}\Big[\beta f(r) \left(2 A r^2+1\right) \left(K r^2+L r^4+1\right) +r^2 \left(2 A-K-L r^2\right)+2 \beta^2 Cr^2 f(r) \left(K r^2+L r^4+1\right) +2 \beta C r^2\Big],~~~~\label{eq42}
\end{eqnarray}
\begin{eqnarray}
&&\hspace{-0.6cm}P_\perp(r,\,\beta)=\frac{1}{2 r \left(K r^2+L r^4+1\right)^2}\Big[2 r \left(A^2 \left(K r^4+L r^6+r^2\right)+A \left(K r^2+2\right)-K-2 L r^2\right)+\beta^2 C r \left(K r^2+L r^4+1\right) \Big(4 f(r) \left(A r^2+1\right) \nonumber\\&&\hspace{0.6cm}\times \left(K r^2+L r^4+1\right)+r \left(6 C r+\Omega(r) K r^2+\Omega(r) L r^4+\Omega(r)\right)\Big)+\beta \Big\{2 r \Big(C \left(2 A \left(K r^4+L r^6+r^2\right)+K r^2+2\right)+A f(r) \nonumber\\&&\hspace{0.6cm}\times\left(A r^2+2\right) \left(K r^2+L r^4+1\right)^2\Big)+\Omega(r) \left(A r^2+1\right) \left(K r^2+L r^4+1\right)^2\Big\}+2 \beta^3 C^2 f(r) r^3 \left(K r^2+L r^4+1\right)^2\Big],~~~~~~~~~~\label{eq43}\\
&&\hspace{-0.6cm}\epsilon(r,\,\beta)=\frac{1}{\left(K r^3+L r^5+r\right)^2}\Big[-\beta (\Omega(r) r+f) \left(K r^2+L r^4+1\right)^2+K^2 r^4+K r^2 \left(2 L r^4+3\right)+L r^4 \left(L r^4+5\right)\big].~~~~~~~~~\label{eq44}
\end{eqnarray}
\end{small}
\end{widetext}
The matching conditions (\ref{eq34})-(\ref{eq36}) for new solution (\ref{eq41})-(\ref{eq44})  
\begin{eqnarray}
&& e^{-\lambda(R)}=\frac{1}{1+KR^2+LR^4}+\beta\,f(R)=1-\frac{2 {M}}{R},~~~\\
&& e^{\nu(R)}=e^{AR^2+B+\beta\,CR^2}=1-\frac{2 {M}}{R}, \\
&& P_r(R)=0, 
\end{eqnarray}
determine the constant $B$, total mass $M$, and integration constant $F$ as 
\begin{eqnarray}
&& B= \ln\Big[1-\frac{2M}{R}\Big]-AR^2-\beta\,CR^2,\\
&& M=M_s-\frac{\beta\,R}{2}\,f(R)
\end{eqnarray}
where $m(R)=M$ is the total mass of the deformed compact object corresponding to energy-momentum tensor $\hat{T}_{ij}$  given by equation (\ref{eq4}). However, we avoid to write expression for $F$ due to lengthy expression. 
\begin{figure*}
\centering
\includegraphics[width=8.7cm,height=6.5cm]{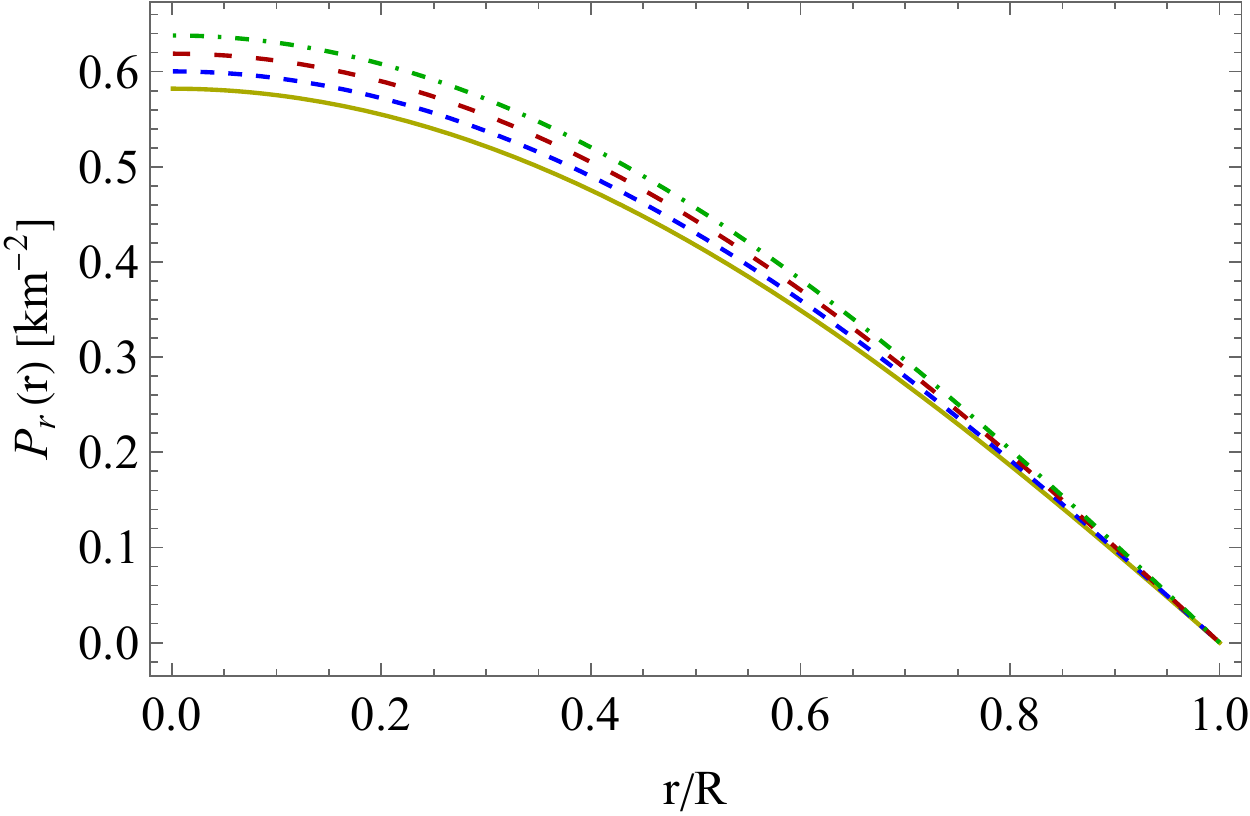}~~~~\includegraphics[width=8.7cm,height=6.5cm]{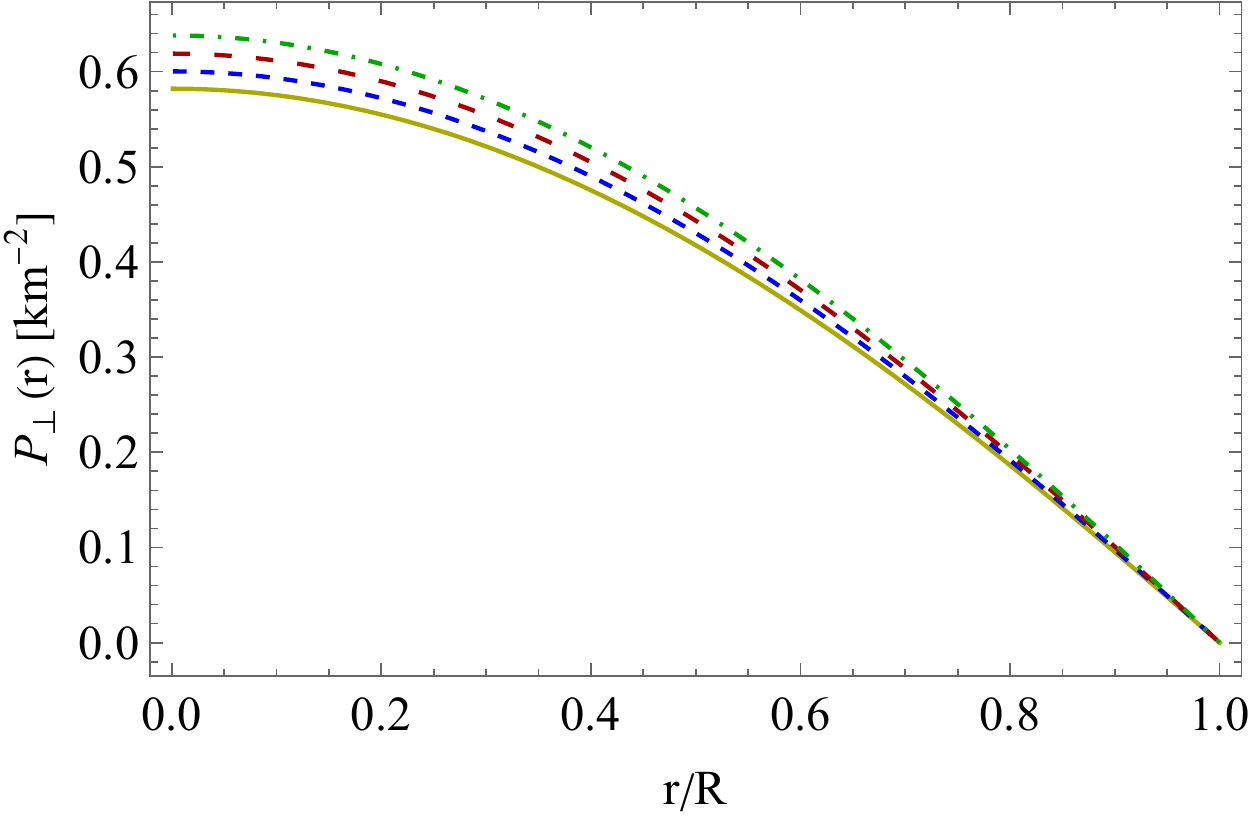}
\includegraphics[width=8.7cm,height=6.5cm]{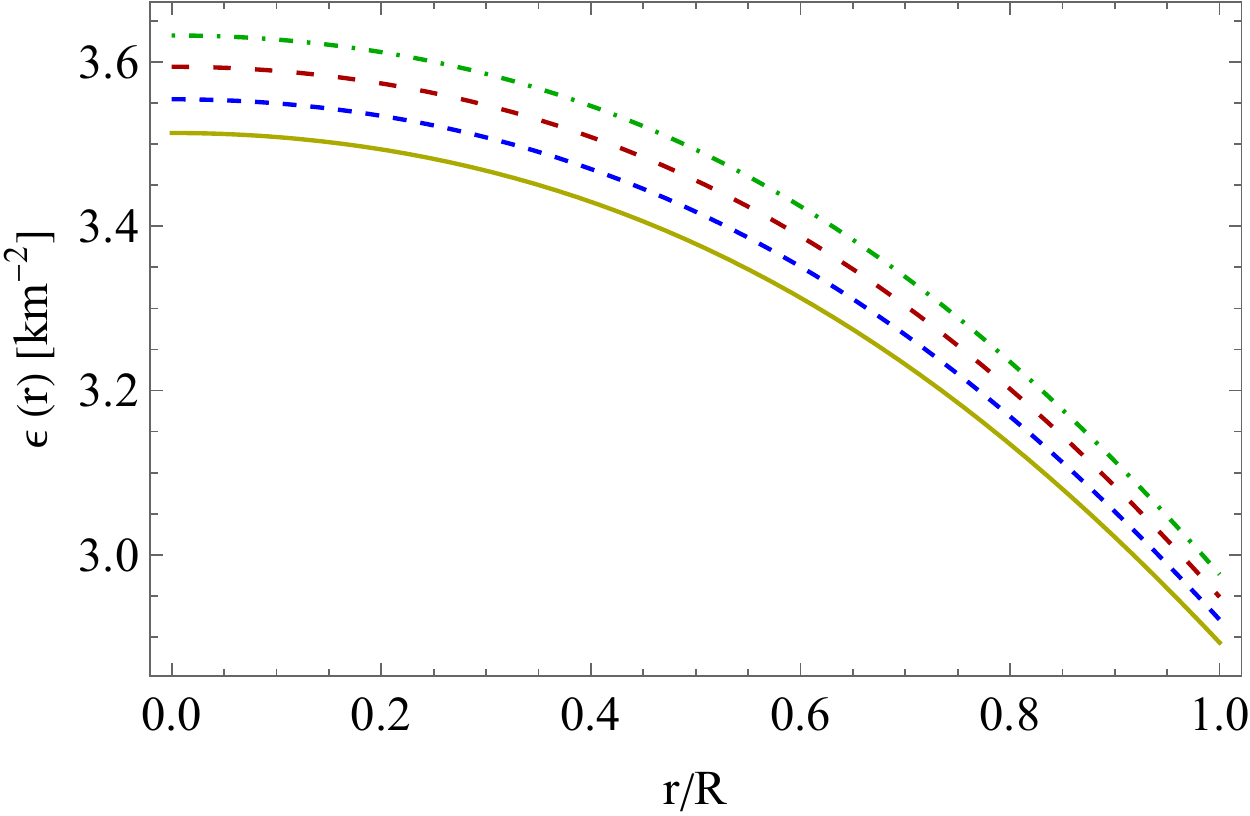}~~~~\includegraphics[width=8.5cm,height=6.5cm]{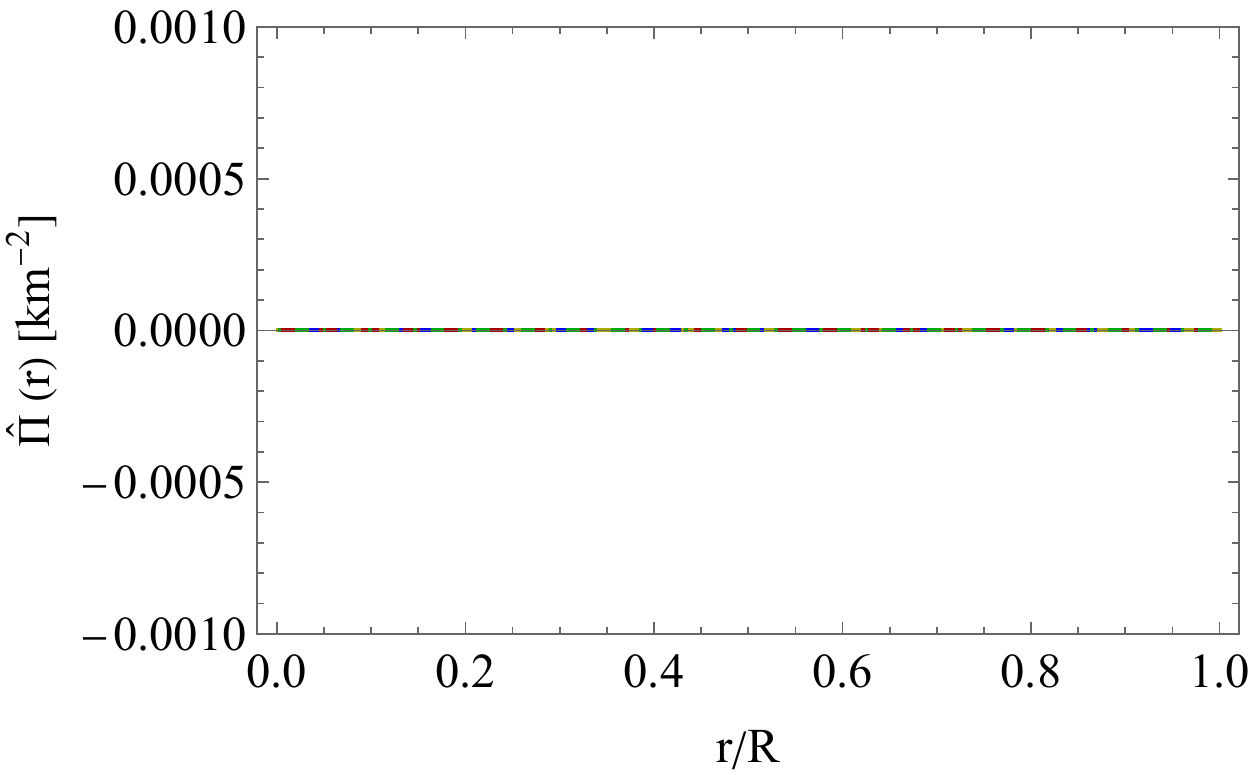}
\caption{The behavior of radial pressure ($P_r\times10^4$)-top left, tangential pressures ($P_\perp\times10^4$)-top right, energy density ($\epsilon\times10^4$)-bottom left and anisotropy ($\hat{\Pi}\times10^4$)-bottom right versus radial coordinate $r/R$ for different coupling constant $\beta$ with compactification factor $\frac{M_s}{R}=0.2$ with $C=0.0002$, and $L= 1.5\times 10^{-8}$. The description of the curves in the figures is as follows: i. Solid-dark yellow for $\beta=0$, ii. dashed-blue for $\beta=0.2$, iii. Dashed-dark red for $\beta=0.4$, and Dotted dash-green for $\beta=0.6$. The figures are plotted corresponding to the isotropic solution discussed in section III.}
\label{f1}
\end{figure*}
It is important to mention here that the expressions (\ref{eq42}) and (\ref{eq43}) given by $P_r$ and $P_\perp$  are the same at each point within the compact object for all $\beta$ i.e. the effective anisotropy $\hat{\Pi}=0$ (see Fig.\ref{f1}), which implies that the solution given by spacetime geometry (\ref{eq41}) represents an isotropic solution of the decoupled system (\ref{eq5})-(\ref{eq7}). Therefore, the gravitational decoupling not only extends the isotropic solution to anisotropic domain but it also plays an important role to convert anisotropic solution in isotropic domain.   

\section{Complexity by gravitational decoupling}
The definition of the complexity factor in static and spherically symmetric self-gravitating systems was initially proposed by Herrera, which is a scalar function denoted by $Y_{TF}$ and it can be measured by anisotropy $\Pi$ and energy density gradient $\rho^\prime$.  Later on Herrara and his collaborators extended this complexity in the context of dynamical spherically symmetric dissipative self-gravitating fluid distributions. Based on the Herrera definition, we denote $\hat{Y}_{TF}$ as a complexity factor for the spherically symmetric static self-gravitating systems (\ref{eq5})-(\ref{eq7}) which is given by, 
\begin{eqnarray} \label{eq50}
\hat{Y}_{TF}= 8\pi \hat{\Pi} -\frac{4\pi}{r^3} \int^r_0 x^3 \epsilon^\prime(x) dx.
\end{eqnarray}
As it is mentioned by Herrera that the complexity factor $\hat{Y}_{TF}$ represents the influence of local anisotropy of
pressure and density inhomogeneity on the Tolman mass ($m_T$) Or, how the Tolman mass is changed by the above two factors defined in $Y_{TF}$. In order to see the influence of $\hat{Y}_{TF}$ on the Tolman $m_T$, we write the Eq.(\ref{eq16}) in terms of complexity factor as,
\begin{eqnarray}
m_T= M_T \Big(\frac{r}{R}\Big)^2+r^3\,\int^R_r \frac{e^{(\nu+\lambda)/2}}{x}\,Y_{TF} dx.  \label{eq51}
\end{eqnarray}
Here, $M_T$ denote the total Tolman mass of the fluid sphere of radius $R$. 

According to Herrera \cite{c22} observations, it is worthwhile noting that \\(i) The complexity factor vanishes for not only isotropic fluid but also for all other configurations where both the terms in (\ref{eq50}) identically vanish.\\
(ii) From the abovementioned criteria, it is evident that there are plenty of configurations with vanishing complexity factors.\\
(iii) It must also be noted that although the contribution of pressure anisotropy to $Y_{TF}$ is local in nature, this is not the case for density energy inhomogeneity.

In the context of MGD,  Casadio et al. argued that the complexity factor satisfies the additive property and then the complexity factor for gravitationally decoupled system will be the sum of two existing complexity factors generated by the sources $T_{ij}$ and $\theta_{ij}$. Therefore, using the above facts,  the complexity factor $\hat{Y}_{TF}$ given by Eq.(\ref{eq50}) can be also written into sum of two complexity factors corresponding to the source $T_{ij}$ and $\Theta_{ij}$ as,
\begin{eqnarray}  \label{eq52}
\hat{Y}_{TF}&=&8\pi\,\hat{\Pi}-\frac{4\pi}{r^3} \int^r_0 x^3 \epsilon^\prime(x) dx,\nonumber\\
&=& 8\pi\,\Pi-\frac{4\pi}{r^3} \int^r_0 x^3 \rho^\prime(x) dx\nonumber\\
&&+ 8\pi\,\beta\,\Pi_{\Theta}-\frac{4\pi\,\beta}{r^3} \int^r_0 x^3 \left[\Theta^0_0(x)\right]^\prime dx, 
\end{eqnarray}
which is denoted as,
\begin{eqnarray}
&& \hat{Y}_{TF}= Y_{TF} + Y^{\Theta}_{TF}.  \label{eq53}
\end{eqnarray}
Here, we denote the $ Y_{TF}$ is the complexity factor for the system (\ref{eq21})-(\ref{eq23}) while $Y^{\Theta}_{TF}$ for (\ref{eq27})-(\ref{eq29}) corresponding to the sources $T_{ij}$ and $\Theta_{ij}$, respectively. \\ 
Now we will discuss two following cases: (A). Complexity factor generated by the isotropic solution (\ref{eq40})-(\ref{eq44}) for the energy-momentum tensor $\hat{T}_{ij}$, and (B). Some new solutions generated EGD approach for the systems having same or vanishing complexity factors.  

\subsection{Complexity factor generated by isotropic solution (\ref{eq41})-(\ref{eq44})}
The complexity factor for the systems (\ref{eq5})-(\ref{eq7}) corresponding to the energy-momentum tensor $\hat{T}_{ij}$ is
\begin{eqnarray}
\hat{Y}_{TF}&=&8\pi\,\hat{\Pi}-\frac{4\pi}{r^3} \int^r_0 x^3 \epsilon^\prime(x) dx, \label{eq54}
\end{eqnarray} 
Since the solution (\ref{eq41})-(\ref{eq44}) is isotropic, then the effective anisotropy will be zero i.e.  $\hat{\Pi}=0$, yields
\begin{eqnarray}
\hat{Y}_{TF}&=&-\frac{4\pi}{r^3} \int^r_0 x^3 \epsilon^\prime(x) dx,\label{eq55}
\end{eqnarray}
Using Eq.(\ref{eq5}), we get
\begin{eqnarray}
\hat{Y}_{TF}&=&\frac{1}{r^2}-\frac{e^{-\lambda}}{r^2}-\frac{\lambda^\prime\,e^{-\lambda}}{2r},
\end{eqnarray}
Now using the solution(\ref{eq41}), we find expression for complexity factor $\hat{Y}_{TF}$ 
\begin{eqnarray}
\hat{Y}_{TF}= \frac{r^2 \left(K^2+2 K L r^2+L \left(L r^4-1\right)\right)}{\left(K r^2+L r^4+1\right)^2}+\frac{\beta}{2r^2}  \big[r\,\Omega(r)\nonumber\\-2 f(r)\big].~~~~
\end{eqnarray}
\begin{figure}
\centering
\includegraphics[width=8.7cm,height=6.5cm]{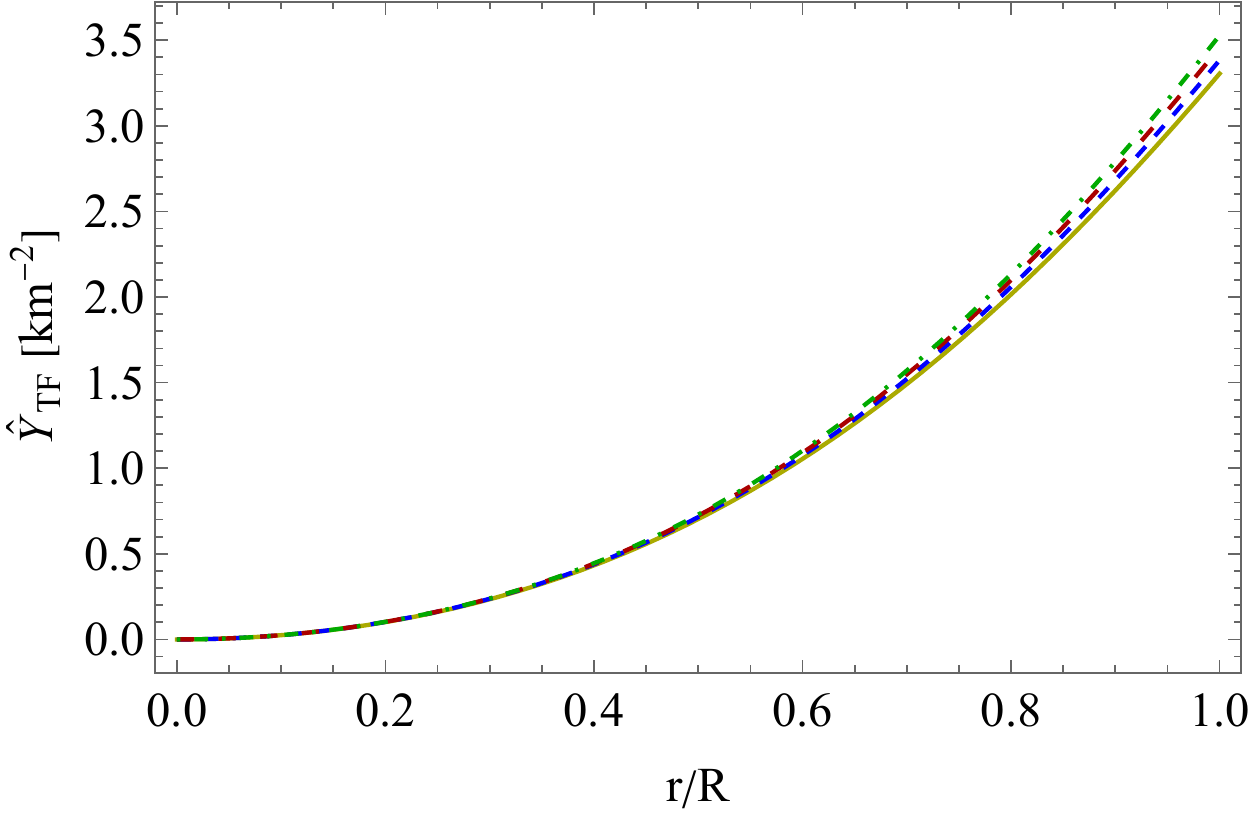}
\caption{The behavior complexity factor $(\hat{Y}_{TF}\times10^4)$ versus radial coordinate $r/R$ for different coupling constant $\beta$ with compactness factor $\frac{M_s}{R}=0.2$ with $C=0.0002$, and $L= 1.5\times 10^{-8}$. The description of the curves in the figures is as follows: i. Solid-dark yellow for $\beta=0$, ii. dashed-blue for $\beta=0.2$, iii. Dashed-dark red for $\beta=0.4$, and Dot dashed-green for $\beta=0.6$. This complexity figure is plotted for the isotropic solution obtained in section III.}  
\label{f2}
\end{figure}
It is noticed from the Fig.\ref{f2}, the decoupling constant $\beta$ is influencing the complexity factor $\hat{Y}_{TF}$. The $\hat{Y}_{TF}$ increases when $\beta$ increase, which implies that gravitational decoupling enhances the complexity of the self-gravitating isotropic models.   

\section{Some new solutions generated EGD approach for the systems having same or vanishing complexity factors}
\subsection{EGD solution for two systems with same complexity factor}
In this section, we will consider the situation where the complexity factor $Y_{TF}$ related to energy-momentum tensor $T_{ij}$ remains same after using gravitational decoupling via CGD, that is $\hat{Y}_{TF}=Y_{TF}$, which implies $Y^{\Theta}_{TF}=0$ or
\begin{eqnarray}
&& \Pi_{\Theta}=\frac{1}{2r^3} \int^r_0 x^3\, \big[ \Theta^0_0(x) \big]^\prime\, dx, \label{eq58}
\end{eqnarray}
where,
\begin{eqnarray}
&& \int^r_0 x^3 \left[\Theta^0_0(x)\right]^\prime dx= \frac{r^3}{4\pi}\Bigg(\frac{f}{r^2}-\frac{f^{\prime}}{2r}\Bigg). \label{eq59}
\end{eqnarray}
Now using the Eqs. (\ref{eq27})-(\ref{eq29}), the equation (\ref{eq58}) yields,
\begin{eqnarray}
 f^{\prime} (4r + \nu^{\prime} r^2)+f (2 \nu^{\prime\prime} r^2 + \nu^{\prime 2} r^2-8 - 2 \nu^{\prime} r) + 
 r (-2 h^{\prime} \mu \nonumber\\+ h^{\prime} \mu^{\prime} r + 2 h^{\prime\prime} \mu r + \beta h^{\prime2} \mu r + 2 h^{\prime} \nu^{\prime} \mu r )=0,~~\label{eq60}
\end{eqnarray}
It is clear that new source $\Theta_{ij}$ can be determined by any solution of the Eq.(\ref{eq60}) which can be obtained through any known solution of the system (\ref{eq21})-(\ref{eq23}) described by the metric functions $\xi$ and $\mu$ together with imposing any viable form of deformation function $f(r)$ or $h(r)$. For this purpose, we consider a well-known spacetime geometry proposed by Karori-Barau,
\begin{eqnarray}
&& e^{\xi(r)}=e^{Ar^2+B}, ~~~~\text{and}~~~\mu(r)=e^{-Dr^2}. \label{eq61}
\end{eqnarray}
where $A$, $B$, and $D$ are constant parameters. Using above $\xi$ and $\mu$, the system (\ref{eq21})-(\ref{eq23}) provides the energy density and pressures expressions for the energy-momentum tensor $T_{ij}$ as,
\begin{eqnarray}
&& p_r=\frac{e^{-D r^2} \left(1-e^{D r^2}+2 A r^2\right)}{r^2}, ~~~\label{eq62}\\
&& p_t=e^{-D r^2} \left[A^2 r^2+A \left(2-D r^2\right)-D\right], ~~~\label{eq63}\\
&& \rho=\frac{e^{-D r^2} \left(-1+e^{D r^2}+2 D r^2\right)}{r^2}. ~~~\label{eq64}
\end{eqnarray}
The constants involved in the solution are determined by the same matching conditions (\ref{eq34})-(\ref{eq36}) for metric functions (\ref{eq61}), which yields
\begin{eqnarray}
&& D=\frac{\ln R-\ln\left[R-2M_{s}\right]}{R^2}, ~~~\label{eq65}\\
&& A=\frac{M_{s}}{(R-2 M_{s}) R^2}, ~~~\label{eq66}\\
&& B=-\frac{M_{s}}{R-2 M_{s}}+\ln\left[1-\frac{2 M_{s}}{R}\right]. ~~~\label{eq67}
\end{eqnarray}
where $m_{GR}(R)=M_s$ is the total mass of the object. Now we find the complexity factor $Y_{TF}$ by using the definition (\ref{eq50}) as,
\begin{eqnarray}
Y_{TF}=\frac{e^{-D r^2} \left(-2+2 e^{D r^2}+A^2 r^4-D r^2 \left(2+A r^2\right)\right)}{r^2},~~~\label{eq68}
\end{eqnarray}
Using the spacetime geometry (\ref{eq61}) together with the same form of temporal deformation function $h(r)=Cr^2$ as used in previous section (\ref{sec3}), we find the radial deformation function by solving of the equation (\ref{eq60}) as,  
\begin{eqnarray}
&&\hspace{-1.5cm}f(r)=-\frac{C (2 A+3 \beta C-D) r^2 f_8(r)}{(A+\beta C)^2\,e^{\frac{2 (A+\beta C-D)}{A+\beta C}+(A+\beta C) r^2}}\nonumber\\&&\hspace{1.5cm}+\frac{\left(2+A r^2+\beta C r^2\right)\, r^2\,F}{e^{(A+\beta C) r^2}\,}, \label{eq69}
\end{eqnarray}
where $F$ is a constant of integration with dimension $length^{-2}$ and
\begin{small}
\begin{eqnarray}
&& f_8(r)=-(A+\beta C) e^{\frac{(A+\beta C-D) \left(2+A r^2+\beta C r^2\right)}{A+\beta C}}+(A+\beta C-D) \nonumber\\ &&\hspace{1cm} \times \left(2+A r^2+\beta C r^2\right) f_{9}(r),\nonumber\\
&& f_9(r)=\text{ExpIntegralEi}\left[\frac{(A+\beta C-D) \left(2+A r^2+\beta C r^2\right)}{A+\beta C}\right],\nonumber
\end{eqnarray}
\end{small}
The deformation function $f(r)$ given in Eq.(\ref{eq69}) leads the same complexity factor $\hat{Y}_{TF}=Y_{TF}$ i.e. $Y^{\Theta}_{TF}=0$ for all $\beta$. Then the deformed metric functions can be read as, 
\begin{eqnarray}
&&\hspace{-0.5cm} e^{-\lambda(r)}=e^{-Dr^2}+\beta\,\Bigg[-\frac{C (2 A+3 \beta C-D) r^2 f_8(r)}{(A+\beta C)^2\,e^{\frac{2 (A+\beta C-D)}{A+\beta C}+(A+\beta C) r^2}}\nonumber\\&&\hspace{1.0cm}+\frac{\left(2+A r^2+\beta C r^2\right)\, r^2\,F}{e^{(A+\beta C) r^2}\,}\Bigg],~~~\label{eq70} \\
&&\hspace{-0.5cm} e^{\nu(r)}=e^{Ar^2+B+\beta\,Cr^2}. ~~~\label{eq71}
\end{eqnarray}
\begin{figure*}
\centering
\includegraphics[width=8.75cm,height=6.5cm]{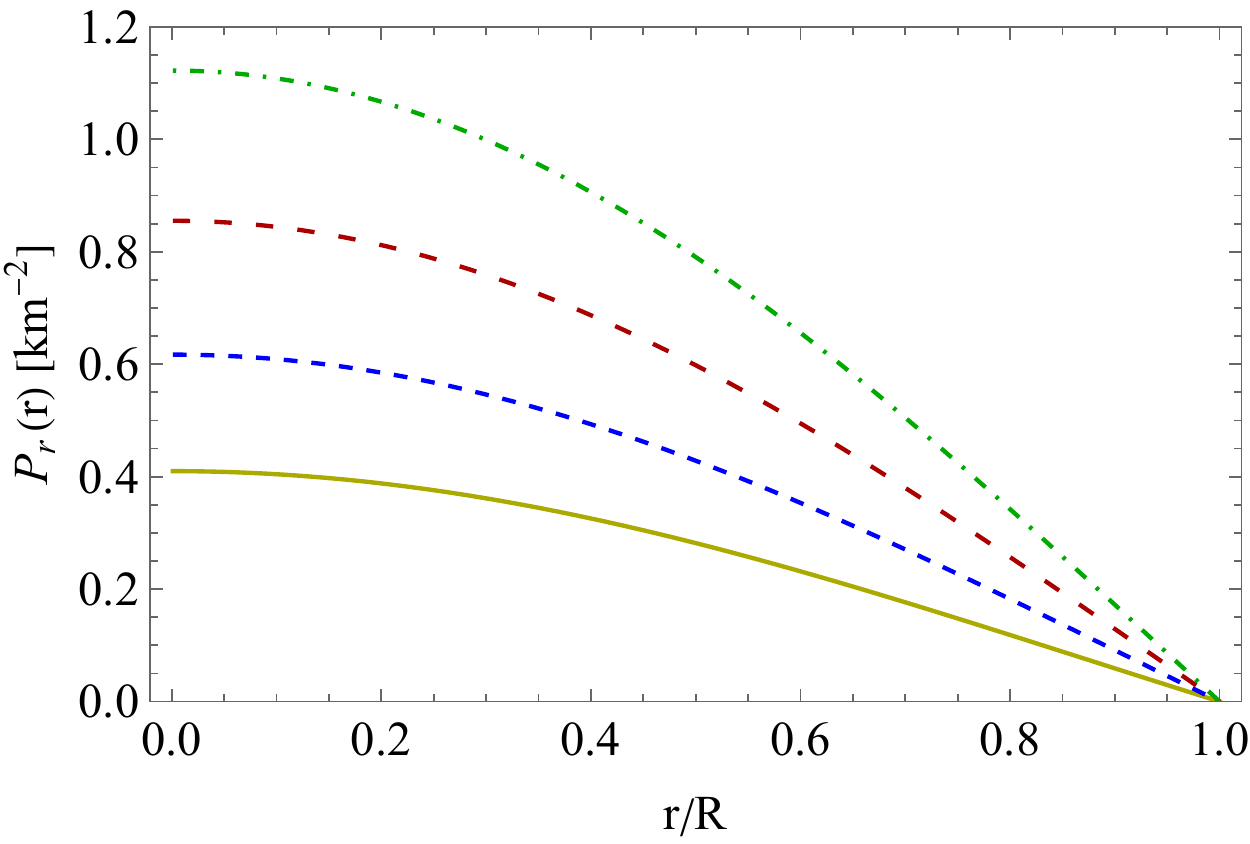}~~~~\includegraphics[width=8.7cm,height=6.5cm]{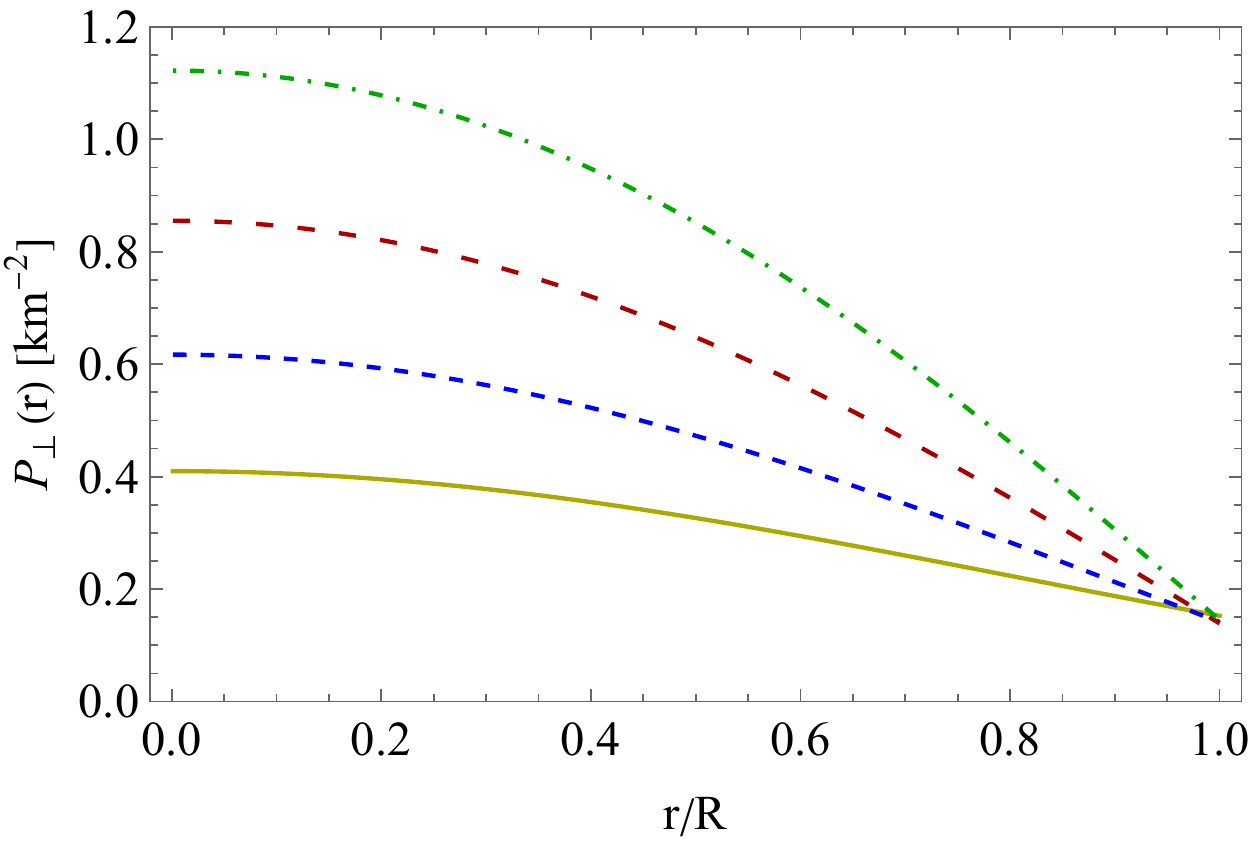}
\includegraphics[width=8.7cm,height=6.5cm]{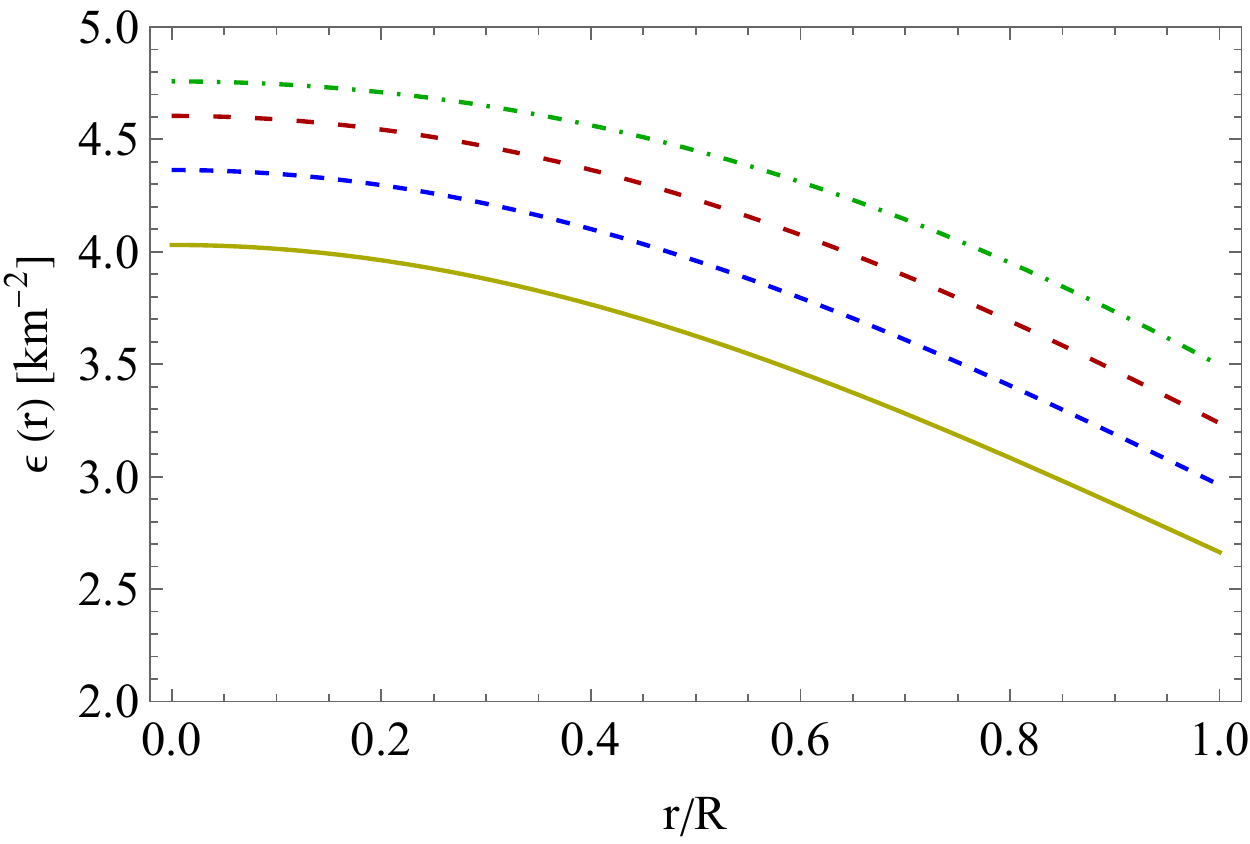}~~~~\includegraphics[width=8.7cm,height=6.5cm]{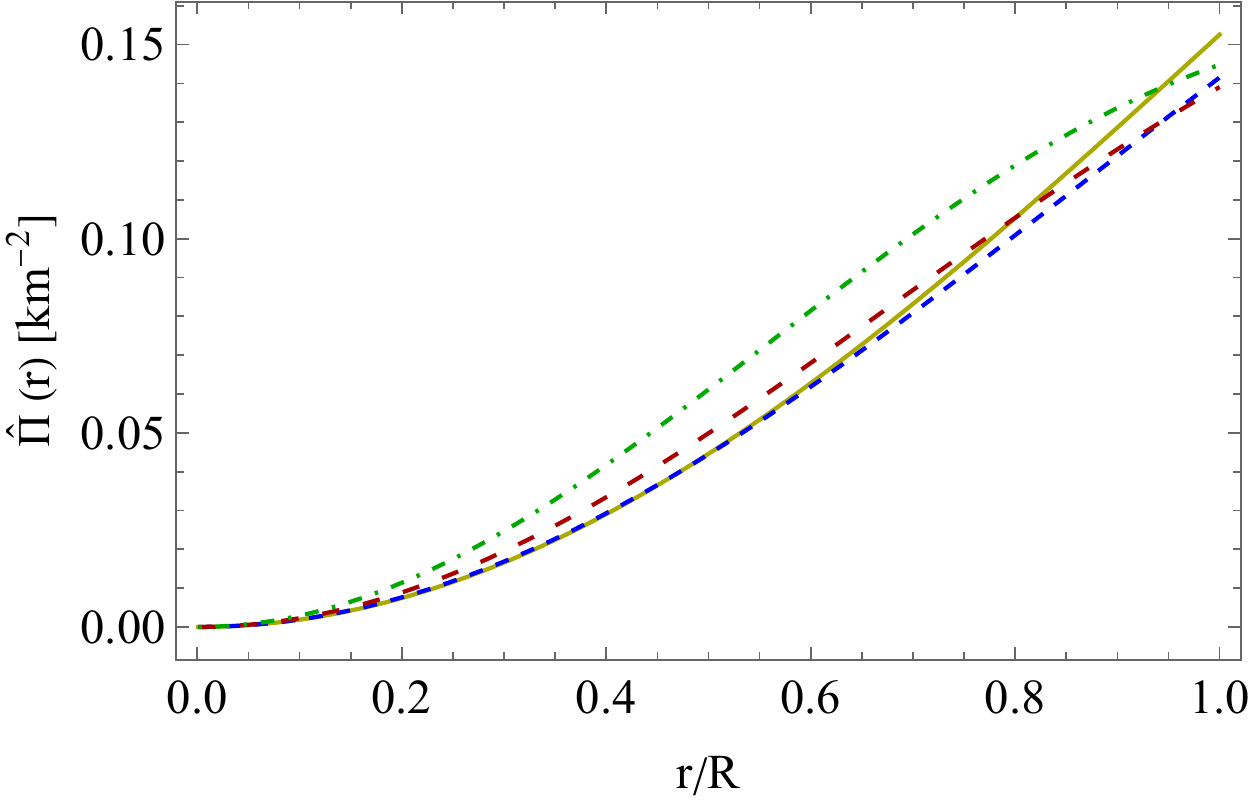}
\caption{The behavior of radial pressure ($P_r\times10^4$)-top left, tangential pressures ($P_\perp\times10^4$)-top right, energy density ($\epsilon\times10^4$)-bottom left and anisotropy ($\hat{\Pi}\times10^4$)-bottom right versus radial coordinate $r/R$ for different coupling constant $\beta$ with compactification factor $\frac{M_s}{R}=0.2$ with $C=0.0002$. The description of the curves in the figures are as follows: i. Solid-dark yellow for $\beta=0$, ii. dashed-blue for $\beta=0.2$, iii. Dashed-dark red for $\beta=0.4$, and Dotted dash-green for $\beta=0.6$. The above figures are plotted for EGD solution for the two systems with same complexity factor obtained in Sec. V (A).}
\label{f3}
\end{figure*}
The expressions for effective pressure and energy density,  
\begin{small}
\begin{eqnarray}
&&\hspace{-0.5cm} P_r(r,\,\beta)=\frac{e^{-D r^2}}{r^2}\, \Big[1+2 A r^2+2 \beta C r^2+e^{D r^2} \big(-1+2 \beta^2 C \nonumber\\&&\hspace{1.0cm}\times f(r) r^2+\beta\,\big\{f(r)+2 A r^2\,f(r)\big\}\big)\Big],~~~\label{eq72}\\
&&\hspace{-0.5cm} P_\perp(r,\,\beta)=\frac{e^{-D r^2}}{r^2} \Big[2+4 \beta C r^2+6 \beta^2 C^2 r^4-2 \beta C D r^4+4 A \nonumber\\&&\hspace{1.0cm}(r^2+\beta C r^4)+e^{D r^2} \big(-2+2 \beta^3 C^2 f(r)\, r^4+\beta^2 C r^2 \nonumber\\&&\hspace{1.0cm} \times \big[r\,\Psi(r)+4 f(r)\, \left(1+A r^2\right)\big]+\beta r [\Psi(r)+A \Psi(r) r^2\nonumber\\&&\hspace{1.0cm} +2 A f(r)\, r \times\left(2+A r^2\right)]\big)\Big], ~~~\label{eq73}\\
&&\hspace{-0.5cm} \epsilon(r,\,\beta)= \frac{e^{-D r^2}}{r^2} \left(2 D\, r^2-1+\frac{ \big(1-\beta\,[f(r)+\Psi(r)\, r]\big)}{e^{-D r^2}}\right), ~~~\label{eq74}
\end{eqnarray}
\end{small}
and effective anisotropy $\hat{\Pi}=P_{\perp}-P_r$ is 
\begin{eqnarray}
&&\hspace{-0.4cm} \hat{\Pi} (r,\,\beta) =\frac{e^{-D r^2}}{2 r^2} \Big[e^{D r^2} \Big(2+2 \beta^3 C^2 f(r)\, r^4+\beta^2 \,C\, r^3 \big(\Psi(r)\,\nonumber\\&&\hspace{1.2cm} +4 A f(r)\, r\big)+\beta \left(1+A r^2\right) \big\{-2 f(r)+\Psi(r) r\nonumber\\&&\hspace{1cm} +2 A  f(r) r^2\big\}\Big) -2 \big(1-A^2 r^4-2 A \beta C r^4-3 \beta^2 C^2\nonumber\\&&\hspace{1cm} \times \, r^4 +D \left(r^2+(A+\beta C) r^4\right)\big)\Big].~~~~~\label{eq75}
\end{eqnarray}
where the expression $\Psi(r)$ is mentioned in the Appendix. The metric functions (\ref{eq70}) and (\ref{eq71}) together with the Eqs.(\ref{eq72})-(\ref{eq75}) represent the complete exact solution of the Einstein field equations (\ref{eq5})-(\ref{eq7}), which is new a anisotropic form of Karori-Barua solution whose complexity factor $\hat{Y}_{TF}$ is same form as the complexity factor $Y_{TF}$ given by Eq.(\ref{eq68}). However, we impose the matching conditions (\ref{eq34})-(\ref{eq36}) under the new solution (\ref{eq69})-(\ref{eq75}) in order to determine the constant parameters $F$ and $B$, and $M$, 
\begin{small}
\begin{eqnarray}
&& \hspace{-0.5cm} F=\Bigg[\frac{\beta C (2 A+3 \beta C-D)\,f_8(R)\, (2 AR^2+2 \beta CR^2+1) }{(A+\beta C)^2e^{\frac{2 (A+\beta C-D)}{A+\beta C}+(A+\beta C) R^2}}\nonumber\\&& \hspace{0.7cm}+\frac{\left(-1+e^{D R^2}-2 A R^2\right)}{R^2e^{D R^2} }-2 \beta C e^{-D R^2}\Bigg] F_{1}(R),\\
&& \hspace{-0.5cm} B= \frac{e^{-DR^2}+\beta\,f(R)}{e^{AR^2+\beta\,CR^2}},\\
&& \hspace{-0.5cm} M=M_s-\frac{\beta\,R}{2}f(R).
\end{eqnarray}
\end{small}
where $M_s=\frac{R}{2}(1-e^{-Dr^2})$, while $F_1(R)$ is mentioned in the Appendix. 
The Fig. \ref{f3} shows the behaviors of  radial and tangential pressures, energy density, and anisotropy inside the self-gravitating anisotropic compact object. it can be observed that all the physical parameters $P_r$, $P_\perp$, and $\epsilon$, and $\hat{\Pi}$ are satisfying the condition for a viable compact object, which implies that the CGD approach is also a very powerful technique to discover new physical viable GD solution for two systems with same complexity factor.\\  The complexity factor $\hat{Y}_{TF}$ takes the form 
\begin{eqnarray}
\hat{Y}_{TF}= \frac{\big[2 e^{D r^2}-2+A^2 r^4-D\, r^2 \left(2+A\, r^2\right)\big]}{r^2\,e^{D r^2}}.
\end{eqnarray}
Here we can generate the family of complexity factors by taking the different values of the compactness factor $\frac{M_s}{R}$. 
\begin{figure}[H]
\centering
\includegraphics[width=8.7cm,height=6.5cm]{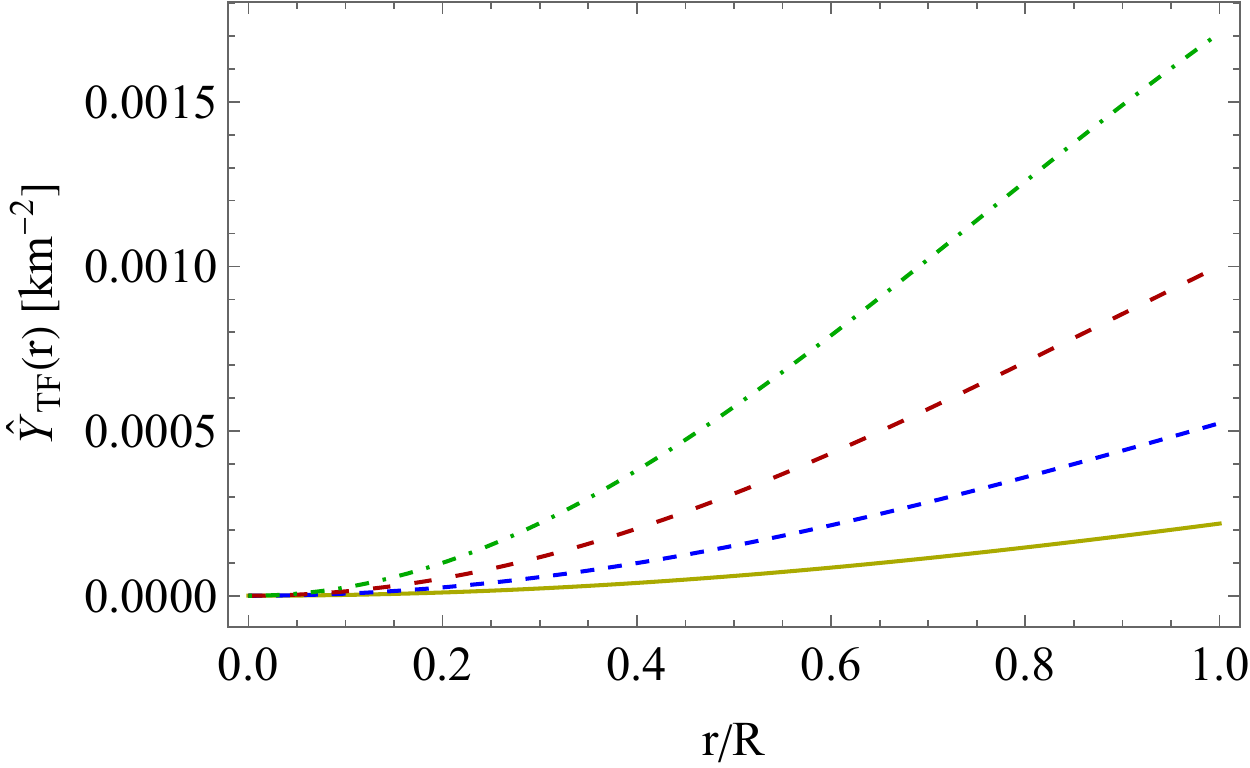}
\caption{The behavior complexity factor $\hat{Y}_{TF}$ versus radial coordinate $r/R$ for different compactification factor $\frac{M_s}{R}$ with $C=0.0002$. The description of the curves in the figures are as follows: i. Solid-dark yellow for $\frac{M_s}{R}=0.10$, ii. dashed-blue for $\frac{M_s}{R}=0.15$, iii. Dashed-dark red for $\frac{M_s}{R}=0.20$, and Dot dashed-green for $\frac{M_s}{R}=0.25$. The above complexity figure is plotted corresponding to solution determined in Sec. V (A).}
\label{f4}
\end{figure}
Since the complexity factor for new anisotropic solution is same as the seed solution. Therefore $\beta$ will not show any direct effect on the complexity. Therefore, we show the influence of the compactness on the complexity factor $\hat{Y}_{TF}$. As we can see from the Fig. \ref{f4}, the complexity is increasing when the compactness factor $\frac{M_s}{R}$ increases.   

\subsection{EGD solution generated by zero complexity factor}
In this section, we discuss the gravitational decoupling solution via complete geometric deformation approach when the complexity factor is zero i.e. $\hat{Y}_{TF}=0$ with the condition $Y_{TF}\ne0$. Therefore, based on the Eq.(\ref{eq52}), we can write 
\begin{eqnarray}
&&\hat{Y}_{TF}= Y_{TF} +Y^{\Theta}_{TF} =0  \nonumber\\&& ~\Longrightarrow~~ Y_{TF} =-8\pi\,\beta\,\Pi_{\Theta}+\frac{4\pi\,\beta}{r^3} \int^r_0 x^3 \Theta^0_0(x) dx, ~~~~\label{eq80}
\end{eqnarray}
Plugging of the Eqs.(\ref{eq27})-(\ref{eq29}) in condition (\ref{eq80}), we determine the following differential equation in geometric deformation functions $f(r)$ and $h(r)$,
\begin{eqnarray}
(4 + \nu^{\prime} r) r\,f^{\prime} - \big[ 2 \nu^{\prime} r+8 r-2 \nu^{\prime\prime}  r^2 - \nu^{\prime 2} r^2 \big] \,f+\big[ r\,\beta h^{\prime 2} \mu r \nonumber\\-2 h^{\prime} \mu r + h^{\prime} \mu^{\prime} r^2 + 2 h^{\prime\prime} \mu r^2 + 2 h^{\prime} \nu^{\prime} \mu r^2 \big]+ \frac{4 r^2}{\beta} Y_{TF}=0,~~\label{eq81}
\end{eqnarray}
Now by considering again the Karori-Barua solution and the complexity factor given by Eqs. (\ref{eq61}) and (\ref{eq68}), respectively together with deformation function $h(r)=Cr^2$, we get the following solution of the above differential equation,
\begin{eqnarray}
&&\hspace{-0.5cm}f(r)=\frac{r^2\,\left(2+A r^2+\beta C r^2\right)}{e^{(A+\beta C) r^2}}   \Bigg[F-\frac{f_{12}(r)}{2 \beta}-f_{13}(r)\Bigg],~~~~~\label{eq82}
\end{eqnarray}
where $F$ is a constant of integration and this above deformation $f(r)$ provides the vanishing complexity factor i.e. $\hat{Y}_{TF}=0$.
The expressions for effective radial and tangential pressures with effective energy density are given as, 
\begin{small}
\begin{eqnarray}
&&\hspace{-0.5cm} P_r(r,\,\beta)=\frac{\left(1-e^{D r^2}+2 A r^2\right)}{r^2\,e^{D r^2}}-\beta \Bigg[\frac{\left(2 Ar^2+2 \beta Cr^2+1\right)}{e^{(A+\beta C) r^2}}\nonumber\\&&\hspace{0.6cm}\times  \left(2+A r^2+\beta C r^2\right) \left(F-\frac{f_{12}(r)}{2 \beta}-f_{13}(r)\right)+\frac{2 C}{e^{D r^2}} \Bigg],~~~~\label{eq83}\\
&&\hspace{-0.5cm} P_\perp(r,\,\beta)=\frac{\left(1-e^{D r^2}+2 A r^2\right)}{r^2\,e^{D r^2}}+\beta\,\Bigg[\frac{\zeta(r) \left(1+A r^2+\beta C r^2\right)}{2 r}\nonumber\\&& \hspace{0.6cm} + \frac{C D \,r^2}{e^{D r^2}}+ \frac{\left(2+2 A r^2+3 \beta C r^2\right)\,C}{e^{D r^2} } + e^{-(A+\beta C) r^2} r^2 \nonumber\\&& \hspace{0.6cm} \times \left(2+A r^2+\beta C r^2\right) \Big\{A^2 r^2+2 A \left(1+\beta C r^2\right)+\beta C \nonumber\\&& \hspace{0.6cm} \times \left(2+\beta C r^2\right)\Big\} \left(F-\frac{f_{12}(r)}{2 \beta}-f_{13}(r)\right)\Bigg], \label{eq84}
\end{eqnarray}
\begin{eqnarray}
&&\hspace{-0.5cm} \epsilon(r,\,\beta)=\frac{\left(e^{D r^2}+2 D r^2-1\right)}{r^2\,e^{D r^2}} - \beta\, \Bigg[\frac{\zeta(r)}{r}+\frac{\left(2+A r^2+b C r^2\right)}{e^{(A+b C) r^2} } \nonumber\\&& \hspace{0.6cm} \times \left(F-\frac{f_{12}(r)}{2 \beta}-f_{13}(r)\right)\Bigg]. \label{eq85}
\end{eqnarray}
\end{small}
where, $\zeta(r)$ is given in the Appendix and the effective anisotropy factor read as,
\begin{small}
\begin{eqnarray}
&& \hspace{-0.5cm} \hat{\Pi}(r,\,\beta)=\frac{e^{-D r^2} \left(-1+e^{D r^2}-2 A r^2\right)}{r^2}+\beta \Bigg(-2 C e^{-D r^2} \nonumber\\&& \hspace{0.6cm} -\frac{\big[2 \beta (F-f_{13}(r))-f_{12}(r)\big] (2 Ar^2+2 \beta Cr^2+1) }{2 \beta e^{(A+\beta C) r^2}(2+A r^2+\beta C r^2)^{-1}}\Bigg)\nonumber\\&& \hspace{0.6cm}-\beta \Bigg[\frac{C Dr^2}{e^{D r^2}}-\frac{\zeta(r) \left(1+A r^2+\beta C r^2\right)}{2 r}- \frac{(A+\beta C)}{e^{(A+\beta C) r^2}}\nonumber\\&& \hspace{0.6cm} \times \frac{\big[2 \beta (F-f_{13}(r))-f_{12}(r)\big] r^2 \left(2+A r^2+\beta C r^2\right)^2}{2 \beta}\nonumber\\&& \hspace{0.6cm}- \frac{C  \left(2+2 A r^2+3 \beta C r^2\right)}{e^{D r^2}}\Bigg]- \frac{\left[D-A^2 r^2+A (D r^2-2)\right]}{e^{D r^2}}.~~~\label{eq86}
\end{eqnarray}
\end{small}
\begin{figure*}
\centering
\includegraphics[width=8.7cm,height=6.5cm]{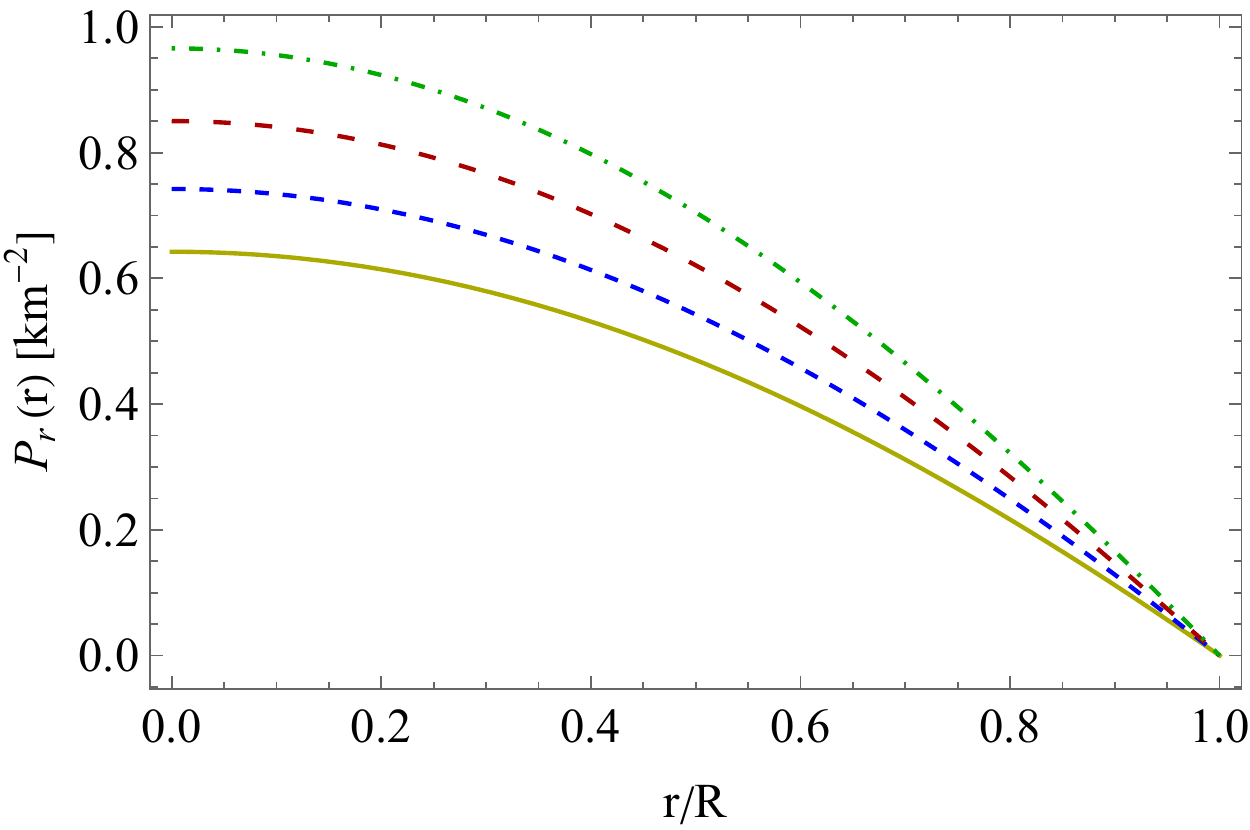}~~~~\includegraphics[width=8.7cm,height=6.5cm]{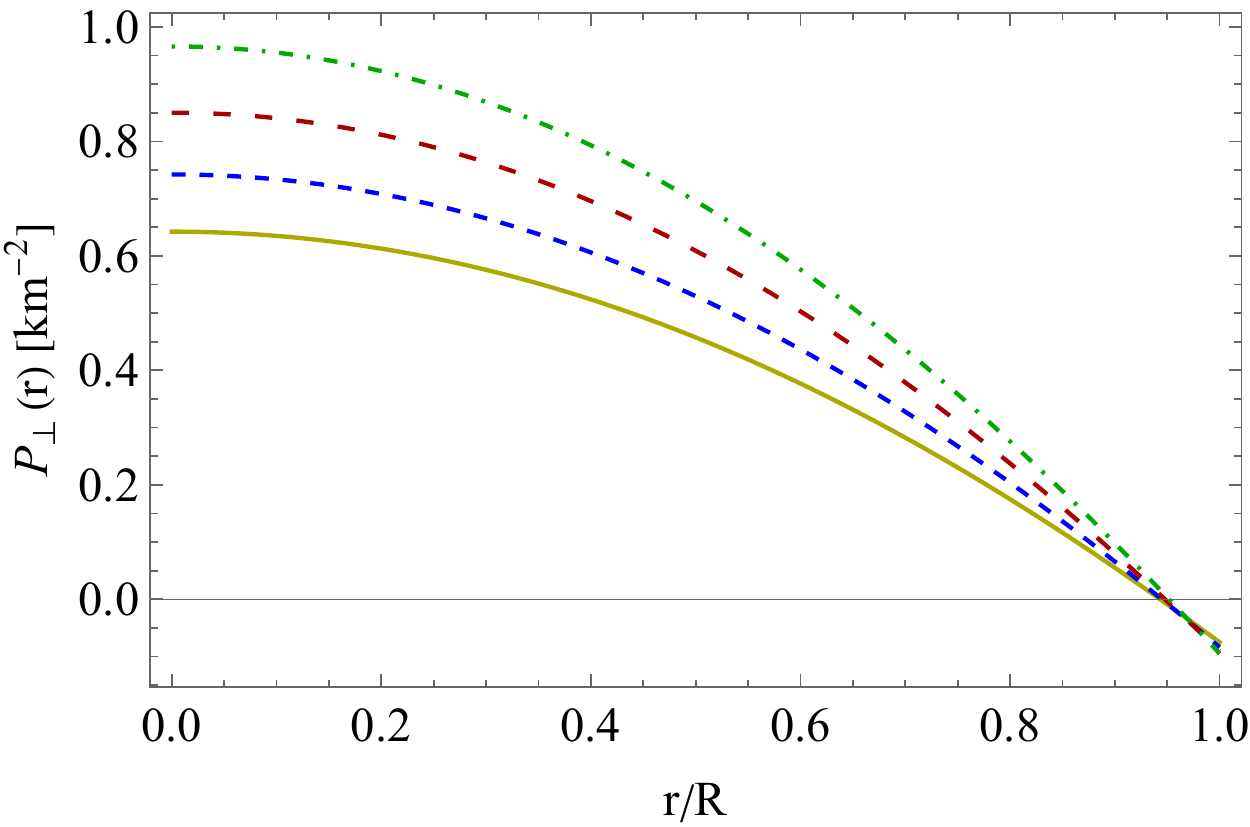}
\includegraphics[width=8.7cm,height=6.5cm]{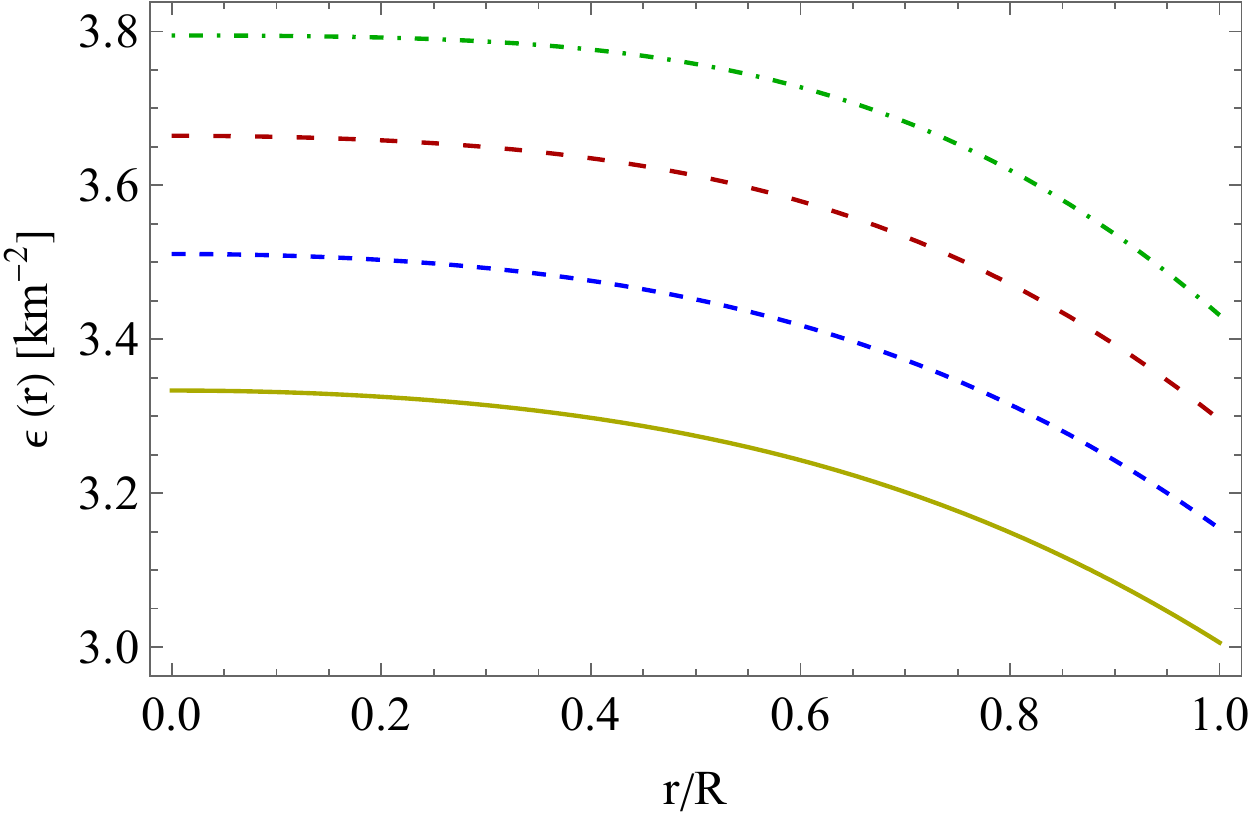}~~~~\includegraphics[width=8.7cm,height=6.5cm]{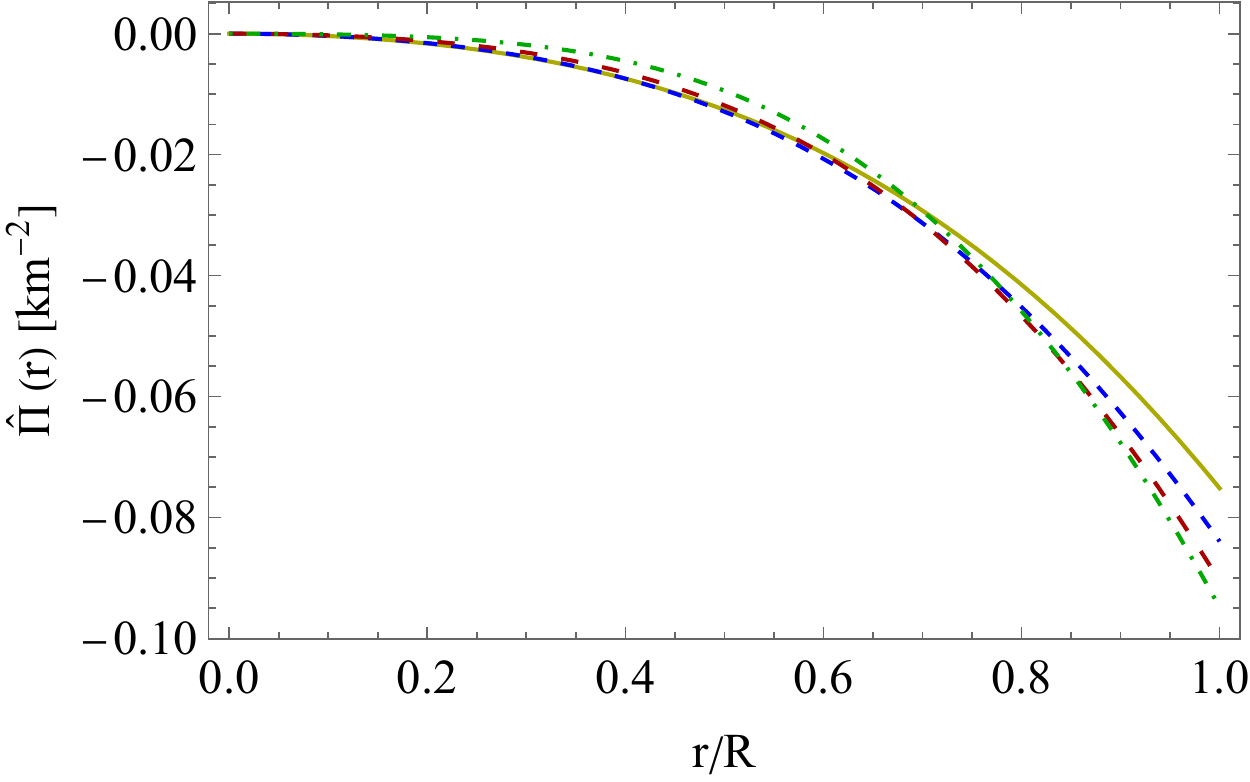}
\caption{The behavior of radial pressure ($P_r\times10^4$)-top left, tangential pressures ($P_\perp\times10^4$)-top right, energy density ($\epsilon\times10^4$)-bottom left and anisotropy ($\hat{\Pi}\times10^4$)-bottom right versus radial coordinate $r/R$ for different coupling constant $\beta$ with compactification factor $\frac{M_s}{R}=0.2$ with $C=0.0002$. The description of the curves in the figures are as follows: i. Solid-dark yellow for $\beta=0$, ii. dashed-blue for $\beta=0.1$, iii. Dashed-dark red for $\beta=0.2$, and Dotted dash-green for $\beta=0.3$. The above figures are plotted corresponding to EGD solution generated by zero complexity
factor presented in in Sec. V (B). }
\label{f5}
\end{figure*}
Now again use the boundary conditions (\ref{eq34})-(\ref{eq36}) for present solution (\ref{eq82})-(\ref{eq86}), we find the $F$, $B$ and total mass $M$,
\begin{small}
\begin{eqnarray}
&& \hspace{-0.5cm} F= \frac{e^{R^2 (A+\beta C)}}{\beta \left(2 A R^2+2 \beta C R^2+1\right) \left(A R^2+\beta C R^2+2\right)} \Bigg[\frac{f_8(R)}{2} \nonumber\\&& \hspace{-0.1cm} \times \frac{\left(A R^2+\beta C R^2+2\right)}{e^{(A+\beta C) R^2}}    \left(2 AR^2+2 \beta CR^2+1\right)+\frac{f_{13}(R)}{2} \nonumber\\&&\hspace{-0.1cm} \times \frac{\left(2 AR^2+2 \beta CR^2+1\right)}{R^2\,e^{\left(R^2 (A+\beta C)\right)+2}} +\frac{ \left(e^{D R^2} -2 A R^2-1\right)}{R^2 e^{D R^2}}-\frac{2 \beta C}{e^{D R^2}} \Bigg],~~~~~ \\
&& \hspace{-0.5cm} B= \frac{e^{-DR^2}+\beta\,f(R)}{e^{AR^2+\beta\,CR^2}},\\
&& \hspace{-0.5cm} M=M_s-\frac{\beta\,R}{2}f(R).
\end{eqnarray}
\end{small}
where $f(R)$ can be determined by Eq.(\ref{eq82}) at $r=R$ while the constant $A$ and $D$ will be same as given by Eqs. (\ref{eq65}) and (\ref{eq66}).
The Fig.\ref{f5} has been plotted against the gravitationally decoupled solution obtained in the context of the zero complexity factor. It is observed that $P_{r}$, $P_\perp$, and $\epsilon$ are monotonically decreasing towards the surface but the tangential pressure ($P_\perp$) is negative near the boundary. This happens due to stronger attractive force generated by the anisotropy, known as anisotropic force ($F_a$), near the surface of the object i.e. $F_a=\frac{2\hat{\Pi}}{r}<0$. The same features also appear under the MGD scenario for Tolman IV solution as discussed by Casadio et al. \cite{m79}. Therefore, we can conclude that the gravitationally decoupled solution under zero complexity factor may not be suitable for modelling of the self-gravitating compact objects.  

\section{Discussions and Conclusions}
In the present article, we have used GD via complete geometric deformation approach to isotropize the self-gravitating anisotropic matter distribution and discussed the complexity of this isotropic solution together with the effect of the decoupling parameter on the complexity. Furthermore, we also investigated two new gravitationally decoupled anisotropic solutions by imposing the condition of two system with same complexity factor as well as systems with zero complexity factor. As we know that most of the previous works, the new solutions were investigated by taking some particular procedures such as equations of state (EoS) for the extra sources added in the original energy-momentum tensor, mimick approaches, and particular ansatz of deformation functions, etc.  
However, in this article we have adopted some different approaches to solve the systems by introducing the gravitational decoupling in the context of CGD. For simplicity, first we started with two energy-momentum tensors in which the first energy-momentum tensor corresponding to anisotropic matter distribution while second one is an unknown source.  As usual, the decoupled system is divided into two sets of equations through the CGD approach by introducing two unknown deformation functions $f(r)$ and $h(r)$ along the radial and temporal component of the metric function of the line element, respectively.  After splitting the field equations, we have considered the following cases:

 In the section III, we have investigated the isotropic solution for the gravitationally decoupled system. For this purpose, first we consider the spactime geometry for seed system corresponding the Tolman-Kuchowicz metric, which is necessary for GD system.  After that we find the isotropic condition of gravitationally decoupled system by employing the effective anisotropy $\hat{\Pi}$ to be zero. In this way we get a differential equation containing two unknowns $f(r)$ and $h(r)$ and solved this equation for $f(r)$ by assuming a particular viable form of $h(r)=Cr^2$. The obtained solution for $f(r)$ gives the vanishing effective anisotropy throughout the star which implies an isotropic solution of the gravitationally decoupled system. 
 
In the section IV, we extended the definition of complexity proposed by Herrera \cite{c22} under gravitational decoupling. This section contains full details about the complexity for the gravitationally decoupled systems. Moreover, we also discussed the complexity factor and impact of decoupling constant $\beta$ on complexity for the obtained isotropic solution in section III. We observe that the complexity is increasing when $\beta$ increases. 

The section V contains the some new solutions generated EGD approach for the systems having same or vanishing complexity factors which is divided into two subsections A and B:\\
A). In the section A, we discover the anisotropic solution by imposing the condition of two systems with same complexity factor by using Krori-Barua seed solution. This said condition leads a vanishing complexity factor for extra source i.e. $Y^{\Theta}_{TF}=0$, which governs a differential equation in $f(r)$ and $g(r)$. This differential equation has been again solved for $f(r)$ by using the same ansatz $h(r)=Cr^2$. Furthermore, we also discussed the complexity factor  and the influence of the compactness on the complexity within the compact objects for the obtained GD anisotropic solution.\\ 
B). The third section C contains the GD anisotropic solution which is obtained by setting zero complexity factor corresponding to the gravitationally decoupled systems i.e. $\hat{Y}_{TF}=0$. The solution obtained in this section shows some drawbacks such that it gives a negative anisotropy throughout the configuration. Since this behavior of the anisotropy leads to an attractive force which may not much suitable for modeling the viable self-gravitating compact objects.  \\
Finally, we would like to mention here that the gravitational decoupling via CGD approach is a very powerful and effective technique to generate new physically viable isotropoic solution from an anisotropic matter distribution. Moreover, this methodology is also useful to generate well-behaved anisotropic solution by assuming of two systems with same complexity factors. In the future projects, we will try to investigate some more conditions on complexity factor for generating the new solutions of the Einstein's field equations for static self-gravitating system by using the gravitational decoupling technique. 

\section*{Acknowledgement}
The author SKM acknowledges that this work is carried out under TRC Project (Grant No. BFP/RGP/CBS-/19/099), the Sultanate of Oman. SKM is thankful for continuous support and encouragement from the administration of University of Nizwa. 

\begin{widetext}
\section*{Appendix}
\begin{small}
\begin{eqnarray}
&&\hspace{-0.5cm}f_1(r)=\text{ExpIntegralEi}[1 + (A+\beta C) r^2],~~~
 f_2(r)=\text{ExpIntegralEi}\left[\frac{(A+\beta C) \left(K-\sqrt{K^2-4 L}+2 L r^2\right)}{2 L}\right],\nonumber
 \end{eqnarray} 
\begin{eqnarray}
 &&\hspace{-0.5cm}f_3(r)=\text{ExpIntegralEi}\left[\frac{(A+\beta C) \left(K+\sqrt{K^2-4 L}+2 L r^2\right)}{2 L}\right],~~\chi_1=\frac{\left(A \left(\sqrt{K^2-4 L}+K\right)+\beta C \left(\sqrt{K^2-4 L}+K\right)-2 L\right)}{e^{-\frac{\left(\sqrt{K^2-4 L}+K\right) (A+\beta C)}{2 L}}}\nonumber,
\\
&&\hspace{-0.5cm} \chi_2=\frac{\left(A \left(\sqrt{K^2-4 L}-K\right)+\beta C \left(\sqrt{K^2-4 L}-K\right)+2 L\right)}{e^{-\frac{\left(K-\sqrt{K^2-4 L}\right) (A+\beta C)}{2 L}}},~~~\chi_3=\sqrt{K^2-4 L} \left(A^2+2 A \beta C-A K+3 \beta^2 C^2-\beta \,C K+L\right)\nonumber,\\
&&\hspace{-0.5cm} \Omega(r)=\frac{r}{e^{\left(r^2 (A+\beta C)\right)}}  \left(\frac{r  \left[2 f_7(r) r \left(K r^2+L r^4+1\right)-2 r\,f_5(r) \left(K+2 L r^2\right)\right]}{\beta\, \chi_3 \left(K r^2+L r^4+1\right)^2\,e^{\frac{A K+\beta C K+L}{L}}}-2 f_4(r) r^2 (A+\beta C)+2 f_4(r)\right)\nonumber\\
&&\hspace{-0.5cm}f_4(r)=F+\frac{1}{{\beta \chi_3 \left(K r^2+L r^4+1\right)}}\Bigg[e^{-\frac{A K+\beta C K+L}{L}} \Bigg(e \Bigg\{\chi_3 \left(K+L r^2\right) e^{\frac{(A+\beta C) \left(K+L r^2\right)}{L}}+\beta^2 C^2 \chi_1 f_2(r) \left(K r^2+L r^4+1\right)+\beta^2 C^2 \nonumber\\
&&\hspace{0.7cm} \times \chi_2 f_3(r) \left(K r^2+L r^4+1\right)\Bigg\}-\chi_3 f_1(r) (A+\beta C) \left(K r^2+L r^4+1\right) e^{\frac{K (A+\beta C)}{L}}\Bigg)\Bigg] \nonumber\\
&&\hspace{-0.5cm} f_5(r)=e \Bigg[\frac{\chi_3 \left(K+L r^2\right)}{e^{-\frac{(A+\beta C)\, (K+L r^2)}{L}}}+\beta^2 C^2 \Big( \chi_1 f_2(r)+ \chi_2 f_3(r) \Big)\left(K r^2+L r^4+1\right)\Bigg]-\frac{\chi_3 f_1(r) (A+\beta C) \left(K r^2+L r^4+1\right)}{ e^{-\frac{K (A+\beta C)}{L}}}\nonumber\\
&&\hspace{-0.5cm} f_6(r)=\frac{\chi_3 L +\chi_3 (A+\beta C) \left(K+L r^2\right)}{e^{-\frac{(A+\beta C) \left(K+L r^2\right)}{L}}} +\beta^2 C^2 \chi_1 f_2(r) \left(K+2 L r^2\right)+\beta^2 C^2 \chi_2 f_3(r) \left(K+2 L r^2\right)+f_{11}(r)+f_{12}(r),\nonumber\\
&&\hspace{-0.5cm}  f_7(r)=-\chi_3\, f_1(r) (A+b C) \left(K+2 L r^2\right) e^{\frac{K (A+b C)}{L}}-\frac{\chi_3 (A+b C)^2 \left(K r^2+L r^4+1\right) e^{\frac{K (A+b C)}{L}+r^2 (A+b C)+1}}{r^2 (A+b C)+1}+e\, f_6(r),\nonumber\\
&&\hspace{-0.5cm}  \Psi(r)=e^{-(A+\beta C) r^2} r\, \bigg[2 (A+\beta C) F r^2+2 F \left(2+A r^2+\beta C r^2\right)-2 (A+\beta C) F\, r^2  \left(2+A r^2+\beta C r^2\right)-\frac{f_{10}(r)+f_{11}(r)}{A+\beta C}-\frac{f_{11}(r)}{(A+\beta C)^2}\bigg],\nonumber\\
&&\hspace{-0.5cm} f_{10}(r)=2 C (A+\beta C-D) (2 A+3 \beta C-D) e^{-\frac{2 (A+\beta C-D)}{A+\beta C}} r^2 f_{9}(r),~~~~ f_{11}(r)=2 C (2 A+3 \beta C-D) e^{-\frac{2 (A+\beta C-D)}{A+\beta C}} r^2  f_{8}(r), \nonumber
\end{eqnarray}
\begin{eqnarray}
&&\hspace{-0.5cm} f_{12}(r)=-(A+\beta C) \text{ExpIntegralEi}\left[(A+\beta C) r^2\right]+\frac{(A+\beta C) \text{ExpIntegralEi}\left[2+A r^2+\beta C r^2\right]}{e^2}+ \frac{2}{(A+\beta C)^2}2 \Bigg[2 \beta^2 C^2 (A+\beta C\nonumber\\
&&\hspace{0.7cm}-D) e^{-\frac{2 (A+\beta C-D)}{A+\beta C}} f_9(r)+\frac{(A+\beta C) e^{(A+\beta C -D) r^2} \left(A+\beta C \left(1-2 \beta C r^2\right)\right)}{r^2 \left[2+A r^2+\beta C r^2\right]}\Bigg], \nonumber,\\
&&\hspace{-0.5cm} f_{13}(r)=\frac{1}{2 \beta e^2 r^2 \left(2+A r^2+\beta C r^2\right)}\bigg(-2 e^{2+A r^2+\beta C r^2} \left(1+A r^2+\beta C r^2\right)+e^2 r^2 \left(A^2 r^2+2 A \left(1+\beta C r^2\right)+\beta C \left(2+\beta C r^2\right)\right) \nonumber\\&&\hspace{0.5cm} \times \text{ExpIntegralEi}\left[(A+\beta C) r^2\right]+r^2 \Big\{A^2 r^2+2 A \left(1+\beta C r^2\right) +\beta C \left(2+\beta C r^2\right)\Big\} \text{ExpIntegralEi}\left[2+(A+\beta C) r^2\right]\bigg),\nonumber\\
&&\hspace{-0.5cm} F_1(R)=\frac{e^{(A+\beta C) R^2}}{\beta (2 AR^2+2 \beta C R^2+1) \left(2+A R^2+\beta C R^2\right)}, ~~~
\zeta(r)=-\frac{1}{\beta (A+\beta C)^2}e^{-2-(2 A+2 \beta C+D) r^2} r (\text{f14}-(A+\beta C) (\text{f15})),\nonumber\\
&&\hspace{-0.5cm} f_{14}(r)=-(A+\beta C)^3 e^{(A+\beta C+D) r^2} \left(-2+A^2 r^4+2 A \beta C r^4+\beta^2 C^2 r^4\right) \text{ExpIntegralEi}\left[2+A r^2+\beta C r^2\right]-4 \beta^2 C^2 (A+\beta C-D) \nonumber\\&&\hspace{0.9cm} \times e^{(A+\beta C) r^2+D \left(\frac{2}{A+\beta C}+r^2\right)} \left(-2+A^2 r^4+2 A \beta C r^4+\beta^2 C^2 r^4\right) \text{f9},\nonumber\\
&&\hspace{-0.5cm}  f_{15}(r)=-2 e^2 \Bigg[\beta^4 C^3 e^{(A+\beta C+D) r^2} F r^4+\beta^3 C^2 r^2 \left(C e^{2 (A+\beta C) r^2} \left(2+e^{D r^2}\right)+3 A e^{(A+\beta C+D) r^2} F r^2\right)+A e^{2 (A+\beta C) r^2} \big(-D+A e^{D r^2} \nonumber\\&&\hspace{0.5cm} \times \left(-1+A r^2\right)\big)+\beta \left(A e^{(A+\beta C+D) r^2} F \left(-2+A^2 r^4\right)-C e^{2 (A+\beta C) r^2} \left(D-A e^{D r^2} \left(-2+3 A r^2\right)\right)\right)+\beta^2 C \Big(e^{(A+\beta C+D) r^2} F \nonumber\\&&\hspace{0.5cm} \times \left(-2+3 A^2 r^4\right)+C e^{2 (A+\beta C) r^2} \left(-2+2 A r^2+e^{D r^2} \left(-1+3 A r^2\right)\right)\Big)\Bigg]+(A+\beta C)^2 e^{(A+\beta C+D) r^2} \Big(-2+A^2 r^4+2 A \beta C r^4 \nonumber\\&&\hspace{0.5cm} +\beta^2 C^2 r^4\Big) \text{ExpIntegralEi}\left[2+(A+\beta C) r^2\right]. \nonumber
\end{eqnarray}
\end{small}
\end{widetext}

\end{document}